\documentclass[10pt,journal,letterpaper,draftclsnofoot,onecolumn]{IEEEtran}
\usepackage[UglyObsolete]{diagrams}
\usepackage{amsmath}
\usepackage{amssymb}
\usepackage[dvips]{graphicx}
\usepackage{latexsym}
\usepackage{stfloats}
\usepackage{color}%\textcolor{red}{
\newtheorem{Main}{Main Lemma}
\newtheorem{definition}{Definition}
\newtheorem{proposition}{Proposition}
\newtheorem{corollary}{Corollary}

\newtheorem{algorithm}{Algorithm}
\newtheorem{example}{Example}
\newtheorem{remark}{Remark}
\renewcommand{\QED}{\QEDopen}
\begin{document}

\title{Lemma for Linear Feedback Shift Registers and DFTs Applied to Affine Variety Codes}

\author{Hajime~Matsui
\thanks{Manuscript submitted October 28, 2012.
This work was supported in part by KAKENHI, Grant-in-Aid for Scientific Research (C) (23560478).
The material in this paper was presented in part at the 34th Symposium on Information Theory and Its Applications, Ousyuku, Iwate, Japan, Nov.~29--Dec.~2, 2011.}% <-this % stops a space
\thanks{H.\ Matsui is with
Toyota Technological Institute,
Hisakata 2--12--1, Tenpaku, Nagoya 468--0034, Japan
(e-mail: matsui@toyota-ti.ac.jp).}
%\thanks{Digital Object Identifier 00.0000/TIT.2005.1200000}}
}

%\markboth{IEEE Transactins on Information Theory,~Vol.~59, No.~4,~April~2013}{Matsui: Lemma for Linear Feedback Shift Registers and DFTs Applied to Affine Variety Codes}

\maketitle

\begin{abstract}
In this paper, we establish a lemma in algebraic coding theory that frequently appears in the encoding and decoding of, e.g., Reed-Solomon codes, algebraic geometry codes, and affine variety codes.
Our lemma corresponds to the non-systematic encoding of affine variety codes, and can be stated by giving a canonical linear map as the composition of an extension through linear feedback shift registers from a Gr\"obner basis and a generalized inverse discrete Fourier transform.
We clarify that our lemma yields the error-value estimation in the fast erasure-and-error decoding of a class of dual affine variety codes.
Moreover, we show that systematic encoding corresponds to a special case of erasure-only decoding.
The lemma enables us to reduce the computational complexity of error-evaluation from $O(n^3)$ using Gaussian elimination to $O(qn^2)$ with some mild conditions on $n$ and $q$, where $n$ is the code length and $q$ is the finite-field size.
\end{abstract}

\IEEEpeerreviewmaketitle

%%%%%%%%%%%%%%%%%%%%%%%%%%%%%%%%%%%%%%%%%%%%%%%%%%%%%%%%%%%%%%%%
\section{Introduction
\label{Introduction}}

Affine variety codes \cite{Fitzgerald-Lax98},\cite{Geil08},\cite{Marcolla-Orsini-Sala},\cite{Miura98} belong to a naturally generalized class of algebraic geometry (AG) codes, and are also known as evaluation codes from order domains of finitely generated types \cite{Andersen-Geil08},\cite{Geil-Pellikaan02},\cite{Little07},\cite{Little09}.
It is known \cite{Fitzgerald-Lax98} that affine variety codes represent all linear codes.
On the other hand, Pellikaan {\it et al.} \cite{Pellikaan-Shen-van Wee91} have already shown that AG codes, especially codes on algebraic curves, also represent all linear codes.
Thus, from the viewpoint of code construction, one might consider only codes on algebraic curves.
However, in terms of decoding, it is insufficient to focus only on AG codes, because many efficient decoding algorithms can correct errors up to half the generalized Feng--Rao minimum distance bound $d_{\mathrm{FR}}$ \cite{generic06},\cite{Matsumoto-Miura2000},\cite{Ruano},\cite{Salazar-Dunn-Graham}, which depends on orders among vector basis or monomial basis.
Whereas AG codes use a specified order, affine variety codes have the advantage that they can choose their orders flexibly, allowing them to reach potentially good $d_{\mathrm{FR}}$ values.

Pellikaan \cite{Pellikaan92} developed a decoding algorithm for all linear codes using a $t$-error correcting pair and solving a system of linear equations; its computational complexity is of the order $n^3$, where $n$ is the code length.
On the other hand, Fitzgerald {\it et al.}~\cite{Fitzgerald-Lax98} and Marcolla {\it et al.}~\cite{Marcolla-Orsini-Sala} proposed decoding algorithms via the Gr\"obner basis that correct errors of up to half the minimum distance $\lfloor(d_\mathrm{min}-1)/2\rfloor$ for affine variety codes.
As this type of decoding belongs to the class of NP-hard problems \cite{Berlekamp-McEliece-van Tilborg78}, it is possible that the algorithms in \cite{Fitzgerald-Lax98},\cite{Marcolla-Orsini-Sala} do not run in polynomial time.

The decoding of dual affine variety codes up to $\lfloor(d_\mathrm{FR}-1)/2\rfloor$ can be divided into two steps, namely error-location and error-evaluation.
For the error-location step, O'Sullivan \cite{generic06},\cite{BMS05} gave a generalization of the Berlekamp--Massey--Sakata (BMS) algorithm for finding the Gr\"obner bases of error-locator ideals for affine variety codes.
The computational complexity of this algorithm is $zn^2$ (where $z$ is the number of elements in the Gr\"obner bases), which is less than $n^3$.
However, for the error-evaluation step in the decoding, no efficient method with a computational complexity of less than $n^3$ has been found.
Although there is a method \cite{Little07} for error-value estimation based on the generalization of the key equation, its relation to the BMS algorithm has not been clarified, as discussed in page 15 of \cite{Little07}, and its computational complexity has not been determined.
Another method that uses the inverse matrix of the proper transform was introduced by Saints {\it et al.} \cite{Saints95}, but its computational complexity is of the order $n^3$, because the inverse matrix must be computed for each error-evaluation step per decoding.
Thus, there is currently no efficient method for error-value estimation in conjunction with the BMS algorithm.

The contents of this paper can be divided into three parts.
First, we realize a generalization of the $N$-dimensional ($N$-D) discrete Fourier transform (DFT) and its inverse (IDFT) over finite fields, where $N$ is a positive integer.
Let $q$ be a prime power, $\mathbb{F}_q$ be the finite field of $q$ elements, $\mathbb{F}_q^\times=\mathbb{F}_q\backslash\{0\}$, $\left(\mathbb{F}_q^\times\right)^N$ be the set of all $N$-tuples of elements in $\mathbb{F}_q^\times$, and $\mathbb{F}_q^N$ be similar to $\left(\mathbb{F}_q^\times\right)^N$.
Whereas the conventional $N$-D DFT and IDFT over finite fields are defined upon vectors whose components are indexed by $\left(\mathbb{F}_q^\times\right)^N$, our generalized transforms are defined upon vectors whose components are indexed by $\mathbb{F}_q^N$, and agree with the conventional ones if they are restricted to $\left(\mathbb{F}_q^\times\right)^N$.
In particular, our generalized transforms satisfy the Fourier inversion formulae; the inclusion-exclusion principal plays an essential role in their proofs.

Secondly, we prove a lemma, which we call Main Lemma, concerning the linear feedback shift registers made by Gr\"obner bases and the generalized IDFT.
Intuitively, Main Lemma corresponds to the encoding of affine variety codes with zero-dimensional information.
We reveal that Main Lemma describes not only non-systematic and systematic encoding but also the error-value estimation.
More specifically, Main Lemma provides a canonical isomorphic map from one vector space, consisting of vectors whose components are indexed by $D(\Psi)$, onto another vector space consisting of vectors whose components are indexed by $\Psi$.
Here, for any subset $\Psi$ of $\mathbb{F}_q^N$, $D(\Psi)$ is the delta set (or footprint) of the Gr\"obner basis for an ideal of $N$-variable polynomials over $\mathbb{F}_q$ that have zeros at $\Psi$.
Although these two vector spaces have the same dimension and are obviously isomorphic, our Main Lemma asserts that there is a canonical one-to-one map that does not depend on the choice of the bases of the vector spaces.
This canonical isomorphic map can be explicitly written as the composition of the generalized IDFT after a map coming from the linear recurrence relations given by the Gr\"obner bases.
The inverse of this canonical isomorphic map agrees with the proper transform introduced by Saints {\it et al.} \cite{Saints95}.

Finally, Main Lemma is applied to affine variety codes in the following three topics.
The first is the construction of affine variety codes, specifically their non-systematic encoding.
Usually, the parity check matrices of affine variety codes must be derived from their generator matrices through matrix elimination.
Using Main Lemma, we directly construct the dual affine variety codes as images of the canonical isomorphic map; this is analogous to the direct construction of affine variety codes as the images of the evaluation map.
The second topic is the error-value estimation in the fast erasure-and-error decoding of a class of dual affine variety codes.
We show that there is an efficient error-value estimation in conjunction with the BMS algorithm.
Our method corresponds to a generalization of the methods of Sakata {\it et al.}~\cite{Sakata-Jensen-Hoholdt95},\cite{Sakata-erasure98} for error-value estimation by DFT in case of one-point AG codes from algebraic curves, a subclass of AG codes.
The final topic is the systematic encoding of the class of dual affine variety codes and the improved erasure-correcting capability.
If a linear code has a non-trivial automorphism group, then it can be encoded systematically by the method of Heegard {\it et al.}~\cite{Heegard95} and Little \cite{Little09}.
Our systematic encoding does not use any automorphism group, and is applicable to a sufficiently wide class of dual affine variety codes.
Moreover, we reveal that systematic encoding is a special type of erasure-only decoding; this fact is well-known in the case of maximum-distance codes \cite{Blahut83}, and is shown for the class of dual affine variety codes.

The content of this paper is attributable to the author, except for the definition of proper transforms \cite{Saints95}, the definition of affine variety codes \cite{Fitzgerald-Lax98}, and the error-value estimation of AG codes \cite{Sakata-Jensen-Hoholdt95},\cite{Sakata-erasure98}.
This other content is still the author's work, even for the limited case of AG codes.
Some of the results in this paper have already been presented in \cite{SITA11}.
The crucial advantage of this paper over \cite{SITA11} is that we can apply a multidimensional (m-D) DFT algorithm \cite{Blahut83} to the generalized DFT and IDFT in order to reduce their computational complexities.
Whereas both computational complexities are estimated as of order $nNq^N$ in \cite{SITA11}, we reduce this to order $Nq^{N+1}$; if $n\ge q$, then the order is actually improved.
For practical use, a faster DFT algorithm, which has less complexity, is adopted.
Thus, the results in \cite{SITA11} are sufficiently extended in this paper to enable us to implement it.
On the other hand, if one adopts the conventional $N$-D DFT and IDFT over $\left(\mathbb{F}_q^\times\right)^N$ in place of our generalized transforms, the results in \cite{ISITA12} then correspond to the special cases of our Main Lemma and its application to a subclass of dual affine variety codes.

As mentioned above, because the isomorphic map of Main Lemma is equivalent to the inverse map of the proper transform in \cite{Saints95}, the above applications to affine variety codes can also be performed by multiplying by the inverse matrix of the proper transform.
Nevertheless, our IDFT-based expression of the inverse map enables us to reduce the computational complexity; moreover, this can be reduced further by applying an m-D DFT algorithm or FFT.
Whereas the computational complexity of the error-value estimation with Gaussian elimination is of the order $n^3$, that with the proposed method has an upper bound of the order $nq^N$, which is equivalent to $qn^2$ because $N$ can be chosen as $q^{N-1}<n\le q^N$.
Thus, our generalized IDFT and Main Lemma are not only important in the theory of affine variety codes, but are also useful in reducing the computational complexity of their error-value estimation.

The rest of this paper is organized as follows.
In Section \ref{Notation}, we prepare some notation for the subsequent discussions.
Section \ref{Fourier-type} gives a generalization of DFTs from $\left(\mathbb{F}_q^{\times}\right)^N$ to $\mathbb{F}_q^N$.
In Section \ref{Main lemma}, we state Main Lemma; Subsection \ref{vector spaces} defines two vector spaces via Gr\"obner bases, Subsection \ref{Extension map} defines the map from the linear feedback shift registers given by Gr\"obner bases, and Subsection \ref{Isomorphic map} gives an isomorphism between the two vector spaces.
In Section \ref{Application}, we apply the lemma to construct affine variety codes, reformulate erasure-and-error decoding algorithms, and determine the relation between systematic encoding and erasure-only decoding.
In Section \ref{Estimation}, we estimate the number of finite-field operations in our algorithm; Subsection \ref{Simple} uses a simple count and Subsection \ref{Multidimensional} applies an m-D DFT algorithm.
Section \ref{Conclusion} concludes the paper.

%%%%%%%%%%%%%%%%%%%%%%%%%%%%%%%%%%%%%%%%%%%%%%%%%%%%%%%%%%%%%%%%
\section{Notation
\label{Notation}}
\footnote{A list of main notation is given as Table \ref{Notation list} in appendix.}
Throughout this paper, the following notation is used.
Let $\mathbb{N}_0$ be the set of non-negative integers.
For two sets $A$ and $B$, a set $A\backslash B$ is defined as $\{u\in A\,|\,u\not\in B\}$.
For an arbitrary finite set $S$, the number of elements in $S$ is represented by $|S|$, and let $V_S=\left.\left\{\left(v_s\right)_S\,\right|\,s\in S,\,v_s\in\mathbb{F}_q\right\}$ denote an $|S|$-dimensional vector space over $\mathbb{F}_q$ whose components are indexed by elements of $S$.
\footnote{In \cite{Saints95}, $V_S$ is denoted by $\left(\mathbb{F}_q\right)^S$, and $\left(v_s\right)_{S}\in V_S$ is denoted simply by $c\in\left(\mathbb{F}_q\right)^S$; in this paper, because we must distinguish several types of vectors and indexes, we adopt $\left(v_s\right)_{S}\in V_S$.}
Although a vector $\left(v_s\right)_{S}\in V_S$ is usually denoted as $\left(v_s\right)_{s\in S}$, in this paper we use $\left(v_s\right)_{S}$ in place of $\left(v_s\right)_{s\in S}$ for simplicity.
Because we have to treat different vectors, e.g., $\left(r_s\right)_{S}$, $\left(u_s\right)_{S}$, $\left(v_s\right)_{S}$ in $V_S$, we judge from the index $\left(\cdot_s\right)_{S}$ if they belong to $V_S$.
Unless otherwise noted, for any arbitrary subset $R\subseteq S$, the vector space $V_R$ is considered to be a subspace of $V_S$ given by $V_R=\left.\left\{\left(v_s\right)_S\in V_S\,\right|\,v_s=0\mbox{ for all }s\in S\backslash R\right\}$.
A map $f$ from a set $A$ into a set $B$ is represented by
$$f:A\to B\quad\left[a\mapsto f(a)\right].$$

%%%%%%%%%%%%%%%%%%%%%%%%%%%%%%%%%%%%%%%%%%%%%%%%%%%%%%%%%%%%%%%%
\section{Fourier-Type Transforms on $\mathbb{F}_q^N$
\label{Fourier-type}}

%%%%%%%%%%%%%%%%%%%%%%%%%%%%%%%%%%%%%%%%%%%%%%%%%%%%%%%%%%%%%%%%
\subsection{Definitions
\label{Definitions}}

Let $N$ be a positive integer and let
\begin{align}\label{A}
A&=A_N=\{0,1,\cdots,q-1\}^N=\left\{\left.\underline{a}=\left(a_1,\cdots,a_N\right)\,\right|\,a_1,\cdots,a_N\in\{0,1,\cdots,q-1\}\,\right\},\\
\Omega&=\Omega_N=\mathbb{F}_q^N=\left\{\left.\underline{\omega}=\left(\omega_1,\cdots,\omega_N\right)\,\right|\,\omega_1,\cdots,\omega_N\in\mathbb{F}_q\right\}.\label{omg}
\end{align}
In this section, Fourier-type transforms are defined as maps between vector spaces, both of which have dim $q^N$,
\footnote{In Subsection \ref{Definitions}, the only vector spaces that we will use are $V_A$ and $V_\Omega$, whose vectors are represented by $\left(h_{\underline{a}}\right)_A\in V_A$ and $\left(c_{\underline{\omega}}\right)_{\Omega}\in V_\Omega$, respectively.}
\begin{align*}
V_A&=\left\{\left.\left(h_{\underline{a}}\right)_A\,\right|\,\underline{a}\in A,\,h_{\underline{a}}\in\mathbb{F}_q\right\},\\
V_{\Omega}&=\left\{\left.\left(c_{\underline{\omega}}\right)_{\Omega}\,\right|\,\underline{\omega}\in\Omega,\,c_{\underline{\omega}}\in\mathbb{F}_q\right\}.
\end{align*}

\begin{definition}\label{Generalization}
{\it(Generalization of the m-D DFT over $\mathbb{F}_q$)}
A linear map $\mathcal{F}$ is defined by
\begin{equation}\label{DFT}
\mathcal{F}:V_\Omega\to V_A\quad\left[\left(c_{\underline{\omega}}\right)_\Omega\mapsto\left(\sum_{\underline{\omega}\in\Omega}c_{\underline{\omega}}\underline{\omega}^{\underline{a}}\right)_A\right],
\end{equation}
where $\underline{\omega}^{\underline{a}}=\omega_1^{a_1}\cdots\omega_N^{a_N}$, and $\omega^a$ is considered as the substituted value $\omega^a=\left.x^a\right|_{x=\omega}$, i.e., $\omega^a=$1 for all $\omega\in\mathbb{F}_q$ if $a=0$.
The linear map $\mathcal{F}:V_\Omega\to V_A$ of \eqref{DFT} is called a generalized DFT on $\mathbb{F}_q^N$.
\QED
\end{definition}
Then, $\mathcal{F}$ is actually equal to the compound of ordinary DFTs in $N$ and lower dimensions.

\begin{example}\label{Example1}
Assume $N=1$.
Note that, if $a\not=0$ and $\omega=0$, then $\omega^a=0$ trivially holds.
Thus, $\left(h_a\right)_A=\mathcal{F}\left(\left(c_{\omega}\right)_\Omega\right)\in V_A$ can be directly written as
\begin{equation}\label{F1}
h_a=
\left\{
{\renewcommand\arraystretch{1.5}
\begin{array}{ll}
\sum_{\omega\in\Omega}c_\omega\omega^a=\sum_{\omega\in\Omega,\,\omega\not=0}c_\omega\omega^a & a\not=0\\
\sum_{\omega\in\Omega}c_\omega & a=0.
\end{array}}\right.
\end{equation}

Assume $N=2$.
Then, for each $\underline{a}=\left(a_1,a_2\right)\in A$, $\left(h_{(a_1,a_2)}\right)_A=\mathcal{F}\left(\left(c_{(\omega_1,\omega_2)}\right)_\Omega\right)\in V_A$ can be directly written as
\begin{gather*}
h_{(a_1,a_2)}=
\left\{
{\renewcommand\arraystretch{1.4}
\begin{array}{ll}
\sum_{(\omega_1,\omega_2)\in\Omega}c_{(\omega_1,\omega_2)}\omega_1^{a_1}\omega_2^{a_2}=\sum_{(\omega_1,\omega_2)\in\Omega,\,\omega_1\omega_2\not=0}c_{(\omega_1,\omega_2)}\omega_1^{a_1}\omega_2^{a_2} & a_1a_2\not=0\\
\sum_{(\omega_1,\omega_2)\in\Omega}c_{(\omega_1,\omega_2)}\omega_1^{a_1}=\sum_{(\omega_1,\omega_2)\in\Omega,\,\omega_1\not=0}c_{(\omega_1,\omega_2)}\omega_1^{a_1} & a_1\not=0,\,a_2=0\\
\sum_{(\omega_1,\omega_2)\in\Omega}c_{(\omega_1,\omega_2)}\omega_2^{a_2}=\sum_{(\omega_1,\omega_2)\in\Omega,\,\omega_2\not=0}c_{(\omega_1,\omega_2)}\omega_2^{a_2} & a_1=0,\,a_2\not=0\\
\sum_{(\omega_1,\omega_2)\in\Omega}c_{(\omega_1,\omega_2)}
& a_1=a_2=0.
\end{array}}\right.
\end{gather*}

In general, to write $\mathcal{F}$ directly requires $2^N$ equalities.
\QED
\end{example}

\begin{remark}\label{Remark 1}
If we fix the orders of the elements in $A$ and $\Omega$, then the matrix representation of $\mathcal{F}$ follows easily from \eqref{DFT}.
This matrix is used in Appendix \ref{transpose}.
Although the operation of $\mathcal{F}$ by multiplying this matrix does not always give the minimum computational complexity as shown in Subsection \ref{Multidimensional}, we use this matrix as a demonstration for the simple case of $N=1$ and $q=8$.
Let $\alpha\in\mathbb{F}_8$ be $\alpha^3+\alpha+1=0$.
Then, we have $A=\{0,1,\cdots,7\}$ and $\Omega=\{0,1,\alpha,\alpha^2,\cdots,\alpha^6\}$, and fix these orders.
In this remark, we consider $\left(h_a\right)_A$ and $\left(c_\omega\right)_\Omega$ as row vectors $\left(h_a\right)_A=\left(h_0,h_1,\cdots,h_7\right)$ and $\left(c_{\omega}\right)_\Omega=\left(c_{0},c_{1},c_{\alpha},\cdots,c_{\alpha^6}\right)$.
According to \eqref{F1}, we can determine $\left(h_a\right)_A=\mathcal{F}\left(\left(c_\omega\right)_\Omega\right)$ by
$$
\left(h_a\right)_A=\left(c_{\omega}\right)_\Omega
\left[
{\renewcommand\arraystretch{0.9}
\arraycolsep = 0.8mm
\begin{array}{cccccccc}
1&0&0&0&0&0&0&0\\
1&1&1&1&1&1&1&1\\
1&\alpha^1&\alpha^2&\alpha^3&\alpha^4&\alpha^5&\alpha^6&1\\
1&\alpha^2&\alpha^4&\alpha^6&\alpha^1&\alpha^3&\alpha^5&1\\
1&\alpha^3&\alpha^6&\alpha^2&\alpha^5&\alpha^1&\alpha^4&1\\
1&\alpha^4&\alpha^1&\alpha^5&\alpha^2&\alpha^6&\alpha^3&1\\
1&\alpha^5&\alpha^3&\alpha^1&\alpha^6&\alpha^4&\alpha^2&1\\
1&\alpha^6&\alpha^5&\alpha^4&\alpha^3&\alpha^2&\alpha^1&1
\end{array}}
\right].\;\;\QED
$$
\end{remark}
\vspace{0.5em}

\begin{definition}\label{inverse-Fourier}
{\it(Generalization of the m-D IDFT over $\mathbb{F}_q$)}
For each $\underline{\omega}\in\Omega$, a subset $I=I_{\underline{\omega}}=\left\{i_1\,\cdots,i_m\right\}$ of $\{1,2,\cdots,N\}$ is determined such that $\omega_{i_1}\cdots\omega_{i_m}\not=0$ and $\omega_{i}=0$ for all $i\in\{1,2,\cdots,N\}\backslash I$.
A linear map $\mathcal{F}^{-1}$ is then defined by
\begin{equation}\label{inverse}
\mathcal{F}^{-1}:V_A\to V_\Omega\quad
\left[\left(h_{\underline{a}}\right)_A\mapsto\left(c_{\underline{\omega}}\right)_\Omega\right],
\end{equation}
where
\begin{equation}\label{IDFT}
\begin{split}
&c_{\underline{\omega}}=
\\
&(-1)^m\sum_{l_1,\cdots,l_m=1}^{q-1}
\left\{
\sum_{J\subseteq\{1,2,\cdots,N\}\backslash I}(-1)^{|J|}h_{\underline{i}(I,J)}
\right\}
\omega_{i_1}^{-l_1}\cdots\omega_{i_m}^{-l_m},
\end{split}
\end{equation}
$J$ in the sum runs over all subsets of $\{1,2,\cdots,N\}\backslash I$, and $\underline{i}(I,J)=\left(b_1,\cdots,b_N\right)\in A$ is defined by, for $1\le i\le N$,
\begin{equation*}
b_i=\left\{
\begin{array}{ll}
l_i & i\in I\\
q-1 & i\in J\\
0 & i\in\{1,2,\cdots,N\}\left\backslash\left(I\cup J\right)\right..
\end{array}\right.
\end{equation*}
The linear map $\mathcal{F}^{-1}:V_A\to V_\Omega$ of \eqref{inverse} is called a generalized IDFT on $\mathbb{F}_q^N$.
\QED
\end{definition}
For example, if $\omega_1\cdots\omega_N\not=0$ for $\underline{\omega}=\left(\omega_1,\cdots,\omega_N\right)\in\Omega$, then $I$ is equal to $\{1,2,\cdots,N\}$ and there is only one choice of $J=\emptyset$.
In this case, definition \eqref{IDFT} implies 
$$
c_{\underline{\omega}}=
(-1)^N\sum_{l_1,\cdots,l_N=1}^{q-1}
h_{(l_1,\cdots,l_N)}
\omega_{i_1}^{-l_1}\cdots\omega_{i_N}^{-l_N},
$$
in other words, $\mathcal{F}^{-1}$ agrees with the $N$-D IDFT if $\Omega$ is restricted to $\left(\mathbb{F}_q^\times\right)^N$.
\footnote{For this special case, including a motivating example of Reed--Solomon codes, see \cite{ISITA12}.}
In general, for each $\underline{\omega}\in\Omega$, the value $c_{\underline{\omega}}$ in \eqref{IDFT} is equal to a linear combination of IDFTs whose dimensions do not exceed $N$.

\begin{example}\label{Example2}
Assume $N=1$.
If $\omega\not=0\in\Omega$, then $I=\{1\}\subseteq\{1\}$, $J=\emptyset\subseteq\{1\}\backslash I=\emptyset$, and $\underline{i}(I,J)=l_1=i$.
If $\omega=0\in\Omega$, then $I=\emptyset\subseteq\{1\}$, $J=\emptyset$ or $\{1\}\subseteq\{1\}\backslash I=\{1\}$, and $\underline{i}(I,J)=0,q-1$, respectively.
Thus, $\mathcal{F}^{-1}\left(\left(h_a\right)_A\right)=\left(c_\omega\right)_\Omega\in V_\Omega$ can be directly written as
\begin{equation}\label{F1i}
c_\omega=
\left\{
{\renewcommand\arraystretch{1.5}
\begin{array}{ll}
-\sum_{i=1}^{q-1}h_{i}\omega^{-i} & \omega\not=0\\
h_{0}-h_{q-1} & \omega=0.
\end{array}}\right.
\end{equation}
Assume $N=2$.
For $\underline{\omega}=\left(\omega_1,\omega_2\right)\in\Omega$, e.g., if $\omega_1\omega_2\not=0$, then $I=\{1,2\}\subseteq\{1,2\}$, $J=\emptyset\subseteq\{1,2\}\backslash I=\emptyset$, and $\underline{i}(I,J)=(l_1,l_2)=(i,j)$; if $\omega_1\not=0$ and $\omega_2=0$, then $I=\{1\}\subseteq\{1,2\}$, $J=\emptyset$ or $\{2\}\subseteq\{1,2\}\backslash I=\{2\}$, and $\underline{i}(I,J)=\underline{i}(\{1\},J)=(i,0),(i,q-1)$, respectively.
Thus, $\mathcal{F}^{-1}\left(\left(h_{\underline{a}}\right)_A\right)=\left(c_{(\omega_1,\omega_2)}\right)_\Omega\in V_\Omega$ can be directly written as
\begin{gather*}
c_{(\omega_1,\omega_2)}=
\left\{
{\renewcommand\arraystretch{1.4}
\begin{array}{ll}
\sum_{i,j=1}^{q-1}h_{(i,j)}\omega_1^{-i}\omega_2^{-j} & \omega_1\omega_2\not=0\\
-\sum_{i=1}^{q-1}(h_{(i,0)}-h_{(i,q-1)})\omega_1^{-i} & \omega_1\not=0,\,\omega_2=0\\
-\sum_{j=1}^{q-1}(h_{(0,j)}-h_{(q-1,j)})\omega_2^{-j} & \omega_1=0,\,\omega_2\not=0\\
h_{(0,0)}-h_{(0,q-1)}-h_{(q-1,0)}+h_{(q-1,q-1)}
& \omega_1=\omega_2=0.
\end{array}}\right.
\end{gather*}

In general, the summand in each condition of $\underline{\omega}$ consists of $2^{N-m}$ terms, where $m$ is the number of non-zero components in $\underline{\omega}$.
\footnote{For the case $N=3$ of $\mathcal{F}^{-1}$, see \cite{SITA11}.}
\QED
\end{example}

%%%%%%%%%%%%%%%%%%%%%%%%%%%%%%%%%%%%%%%%%%%%%%%%%%%%%%%%%%%%%%%%
\subsection{Properties
\label{Properties}}

\begin{proposition}\label{Fourier}
{\it(Generalization of the Fourier inversion formulae)}
Two linear maps $\mathcal{F}:V_\Omega\to V_A$ and $\mathcal{F}^{-1}:V_A\to V_\Omega$ are the inverse of each other, i.e., $\mathcal{F}^{-1}\left(\mathcal{F}\left(\left(c_{\underline{\omega}}\right)_\Omega\right)\right)=\left(c_{\underline{\omega}}\right)_\Omega$ and $\mathcal{F}\left(\mathcal{F}^{-1}\left(\left(h_{\underline{a}}\right)_A\right)\right)=\left(h_{\underline{a}}\right)_A$.
\QED
\end{proposition}
The proof is described in Appendix \ref{Proof of Fourier}.
This proposition corresponds to one of the basic concepts in this paper.

\begin{remark}\label{Remark 2}
(Continued from Remark \ref{Remark 1})
According to \eqref{F1i}, we can determine $\left(c_\omega\right)_\Omega=\mathcal{F}^{-1}\left(\left(h_a\right)_A\right)$ by
$$
\left(c_{\omega}\right)_\Omega=\left(h_a\right)_A
\left[
\renewcommand\arraystretch{0.9}
\arraycolsep = 0.8mm
\begin{array}{cccccccc}
1&0&0&0&0&0&0&0\\
0&1&\alpha^6&\alpha^5&\alpha^4&\alpha^3&\alpha^2&\alpha^1\\
0&1&\alpha^5&\alpha^3&\alpha^1&\alpha^6&\alpha^4&\alpha^2\\
0&1&\alpha^4&\alpha^1&\alpha^5&\alpha^2&\alpha^6&\alpha^3\\
0&1&\alpha^3&\alpha^6&\alpha^2&\alpha^5&\alpha^1&\alpha^4\\
0&1&\alpha^2&\alpha^4&\alpha^6&\alpha^1&\alpha^3&\alpha^5\\
0&1&\alpha^1&\alpha^2&\alpha^3&\alpha^4&\alpha^5&\alpha^6\\
1&1&1&1&1&1&1&1
\end{array}
\right].
$$
As a consequence of Proposition \ref{Fourier}, this matrix is equal to the inverse of the matrix that appeared in Remark \ref{Remark 1}; actually, this can be directly checked.
\QED
\end{remark}

The next property is dimensional induction.
The results obtained in the rest of this subsection are not used before Subsection \ref{Multidimensional}.
As observed in Examples \ref{Example1} and \ref{Example2}, computing the values $\mathcal{F}\left(\left(c_{\underline{\omega}}\right)_{\Omega}\right)$ and $\mathcal{F}^{-1}\left(\left(h_{\underline{a}}\right)_{A}\right)$ according to their definitions is not simple.
We will show in Section \ref{Estimation} that computational complexities of these values are of the order $Nq^{2N}$, which is significantly high among the other computational procedures.
On the other hand, the m-D DFT algorithm \cite{Blahut83} is applied to conventional DFT and IDFT over $\left(\mathbb{F}_q^\times\right)^N$ and their complexities are reduced.
We now show that the algorithm can also be applied to our generalized DFT and IDFT and that they can be computed inductively from low-dimensional DFTs and IDFTs.
In the rest of this subsection, we index $\mathcal{F}_N=\mathcal{F}:V_{\Omega_N}\to V_{A_N}$ and $\mathcal{F}_N^{-1}=\mathcal{F}^{-1}:V_{A_N}\to V_{\Omega_N}$ because we will treat generalized DFTs and IDFTs in different dimensions.

\begin{proposition}\label{low-dim}
{\it(Reduction to low dimensional DFTs)}
If $N\ge2$, then the generalized DFT $\mathcal{F}_{N}$ in \eqref{DFT} can be computed from $\mathcal{F}_1$ and $\mathcal{F}_{N-1}$ as
\begin{align}\label{FandF}
&\mathcal{F}_{N}\left(\left(c_{\underline{\omega},\omega_{N}}\right)_{\Omega_{N}}\right)=\left(h_{\underline{a},a_{N}}\right)_{A_{N}},\mbox{ where}\\\label{qN}
&\left(h_{\underline{a},a_{N}}\right)_{A_1}=\mathcal{F}_1\left(\left(v_{\underline{a},\omega_{N}}\right)_{\Omega_1}\right)\mbox{ for all }\underline{a}\in A_{N-1},\\
&\left(v_{\underline{a},\omega_{N}}\right)_{A_{N-1}}=\mathcal{F}_{N-1}\left(\left(c_{\underline{\omega},\omega_{N}}\right)_{\Omega_{N-1}}\right)\mbox{ for all }\omega_{N}\in\Omega_1,\label{q1}
\end{align}
and, from $\left(c_{\underline{\omega},\omega_{N}}\right)_{\Omega_{N}}$, we define $\left(c_{\underline{\omega},\omega_{N}}\right)_{\Omega_{N-1}}$ by $c_{\underline{\omega},\omega_{N}}=c_{\left(\omega_1,\cdots,\omega_{N-1},\omega_{N}\right)}$ with $\underline{\omega}=\left(\omega_1,\cdots,\omega_{N-1}\right)\in\Omega_{N-1}$ and fixed $\omega_{N}\in\Omega_1$.
\QED
\end{proposition}

The proof of this proposition is immediately obtained from Definition \ref{Generalization}; we give an additional explanation in Subsection \ref{Multidimensional}.
Note that \eqref{FandF} is not the usual composition map of $\mathcal{F}_1$ and $\mathcal{F}_{N-1}$; however, \eqref{qN} and \eqref{q1} mean to calculate $\mathcal{F}_1$ $q^{N-1}$-times after calculating $\mathcal{F}_{N-1}$ $q$-times.
It is also noted that, if $N\ge2$, then there are many ways to decompose $\mathcal{F}_{N}$ into the lower dimensional DFTs.
Applying Proposition \ref{low-dim} repeatedly, it is possible to compute $\mathcal{F}_{N}$ only by using $\mathcal{F}_1$, and achieve the least computational complexity as shown in Subsection \ref{Multidimensional}.

\begin{proposition}\label{lower-dim}
{\it(Reduction to low dimensional IDFTs)}
If $N\ge2$, then the generalized IDFT $\mathcal{F}_{N}^{-1}$ in \eqref{inverse} can be computed from $\mathcal{F}_1^{-1}$ and $\mathcal{F}_{N-1}^{-1}$ as
\begin{align}
&\mathcal{F}_{N}^{-1}\left(\left(h_{\underline{a},a_N}\right)_{A_{N}}\right)=\left(c_{\underline{\omega},\omega_N}\right)_{\Omega_{N}},\mbox{ where}\\
&\left(c_{\underline{\omega},\omega_N}\right)_{\Omega_1}=\mathcal{F}_1^{-1}\left(\left(v_{\underline{\omega},a_N}\right)_{A_1}\right)\mbox{ for all }\underline{\omega}\in\Omega_{N-1},\\
&\left(v_{\underline{\omega},a_N}\right)_{\Omega_{N-1}}=\mathcal{F}_{N-1}^{-1}\left(\left(h_{\underline{a},a_N}\right)_{A_{N-1}}\right)\mbox{ for all }a_N\in A_1,
\label{F-1andF}
\end{align}
and, from $\left(h_{\underline{a},a_N}\right)_{A_{N}}$, we define $\left(h_{\underline{a},a_N}\right)_{A_{N-1}}$ by $h_{\underline{a},a_N}=h_{\left(a_1,\cdots,a_{N-1},a_N\right)}$ with $\underline{a}=\left(a_1,\cdots,a_{N-1}\right)\in A_{N-1}$ and fixed $a_{N}\in A_1$.
\QED
\end{proposition}
The proof of this proposition is obvious because of Proposition \ref{Fourier}.
It is remarkable that $\mathcal{F}^{-1}$ and $\mathcal{F}$ have the same inductive expressions even though the summand of $\mathcal{F}^{-1}$ is more complex than that of $\mathcal{F}$.
Similar to $\mathcal{F}_{N}$, the complexity computing $\mathcal{F}_{N}^{-1}$ is minimized by using only $\mathcal{F}_1^{-1}$ in all available methods; these estimations will be performed in Section \ref{Estimation}.
As shown in Fig.\ \ref{multidimensional}, if we represent $\left(h_{\underline{a}}\right)_A$ two-dimensionally according to $A=\{0,1,\cdots,q-2\}^2$, then Proposition \ref{lower-dim} in case $N=1$ insists that $\mathcal{F}_2^{-1}$ is decomposed to the vertical operation of $\mathcal{F}_1^{-1}$ for all $a_1$ and the horizontal operation of $\mathcal{F}_1^{-1}$ for all $a_2$, where $(a_1,a_2)\in A_2$ and the resulting value is trivially independent of the order of two operations.
\begin{example}\label{multi}
\begin{figure}[t!]% Fig. 1
\centering
  \resizebox{6.3cm}{!}{\includegraphics{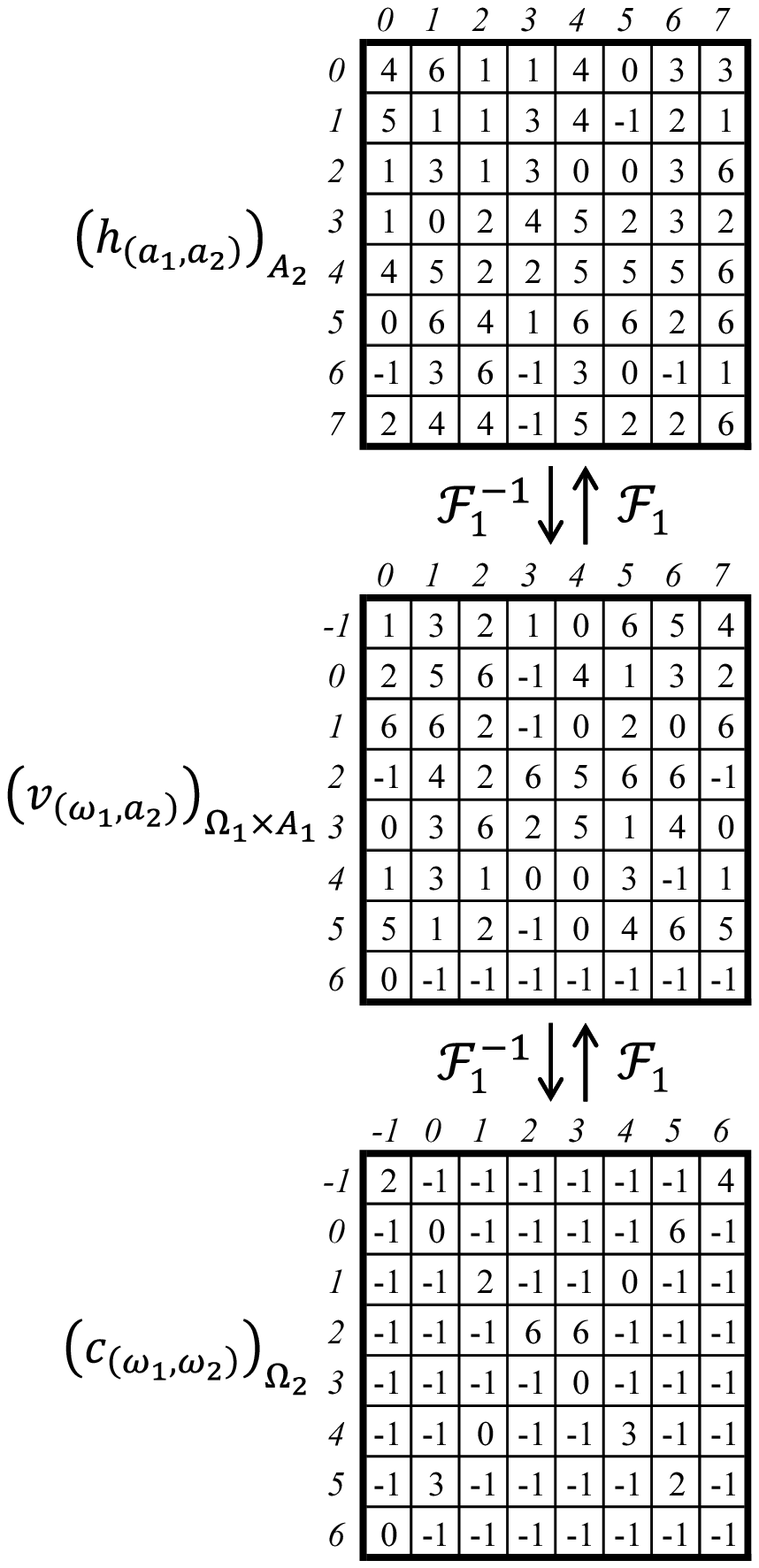}}
\caption{
Numerical example of the m-D DFT algorithm; the details are described in Example \ref{multi}.
The non-zero elements of $\mathbb{F}_8$ are represented by the number of powers of a primitive element $\alpha$ with $\alpha^3+\alpha+1=0$, i.e., $0,1,\cdots,6$ means $\alpha^0,\alpha^1,\cdots,\alpha^6$, respectively, and $-1$ means $0\in\mathbb{F}_8$.
The first $\mathcal{F}_{1}^{-1}$ and $\mathcal{F}_{1}$ mean that the generalized IDFT and DFT are performed for all columns, and the second ones mean that they are performed for all rows.
As a result, $\mathcal{F}_{2}^{-1}\left(\left(h_{(a_1,a_2)}\right)_{A_2}\right)=\left(c_{(\omega_1,\omega_2)}\right)_{\Omega_2}$ and $\left(h_{(a_1,a_2)}\right)_{A_2}=\mathcal{F}_{2}\left(\left(c_{(\omega_1,\omega_2)}\right)_{\Omega_2}\right)$ are verified.
\label{multidimensional}}
\end{figure}
For a given $\left(h_{(a_1,a_2)}\right)_{A_2}$, all values of the m-D DFT algorithm are shown in Fig.\ \ref{multidimensional}, where the vertical and the horizontal axes ({\it 0,1,$\cdots$,7}) in $\left(h_{(a_1,a_2)}\right)_{A_2}$ denote $a_1$ and $a_2$ of $\underline{a}=(a_1,a_2)\in A_2$, the vertical axis ({\it -1,0,$\cdots$,6}) and the horizontal axis ({\it 0,1,$\cdots$,7}) in $\left(v_{(\omega_1,a_2)}\right)_{\Omega_1\times A_1}$ denote $\omega_1$ and $a_2$ of $(\omega_1,a_2)\in\Omega_1\times A_1$, and those axes ({\it -1,0,$\cdots$,6}) in $\left(c_{(\omega_1,\omega_2)}\right)_{\Omega_2}$ denote $\omega_1$ and $\omega_2$ of $\underline{\omega}=(\omega_1,\omega_2)\in\Omega_2$.
At the first step, the first three values of $\mathcal{F}_1^{-1}\left(\left(h_{(a_1,a_2)}\right)_{A_1}\right)=\left(v_{(\omega_1,a_2)}\right)_{\Omega_1}$ at $a_2=0$ are obtained as
\begin{align*}
v_{(-1,0)}&=h_{(0,0)}+h_{(7,0)}=\alpha^4+\alpha^2=\alpha,\\
v_{(0,0)}&=h_{(1,0)}+h_{(2,0)}+h_{(3,0)}+h_{(4,0)}+h_{(5,0)}+h_{(6,0)}+h_{(7,0)}\\
&=\alpha^5+\alpha+\alpha+\alpha^4+1+0+\alpha^2
=\alpha^2,\\
v_{(1,0)}&=h_{(1,0)}\alpha^6+h_{(2,0)}\alpha^5+h_{(3,0)}\alpha^4+h_{(4,0)}\alpha^3+h_{(5,0)}\alpha^2+h_{(6,0)}\alpha+h_{(7,0)}\\
&=\alpha^5\alpha^6+\alpha\alpha^5+\alpha\alpha^4+\alpha^4\alpha^3+\alpha^2+0+\alpha^2
=\alpha^6.
\end{align*}
At the second step, the first three values of $\mathcal{F}_1^{-1}\left(\left(v_{(\omega_1,a_2)}\right)_{A_1}\right)=\left(c_{(\omega_1,\omega_2)}\right)_{\Omega_1}$ at $\omega_1=0$ are obtained as
\begin{align*}
c_{(-1,-1)}&=v_{(-1,0)}+v_{(-1,7)}=\alpha+\alpha^4=\alpha^2,\\
c_{(-1,0)}&=v_{(-1,1)}+v_{(-1,2)}+v_{(-1,3)}+v_{(-1,4)}+v_{(-1,5)}+v_{(-1,6)}+v_{(-1,7)}\\
&=\alpha^3+\alpha^2+\alpha+1+\alpha^6+\alpha^5+\alpha^4
=0,\\
c_{(-1,1)}&=v_{(-1,1)}\alpha^6+v_{(-1,2)}\alpha^5+v_{(-1,3)}\alpha^4+v_{(-1,4)}\alpha^3+v_{(-1,5)}\alpha^2+v_{(-1,6)}\alpha+v_{(-1,7)}\\
&=\alpha^3\alpha^6+\alpha^2\alpha^5+\alpha\alpha^4+\alpha^3+\alpha^6\alpha^2+\alpha^5\alpha+\alpha^4
=0.
\end{align*}
In this numerical example, we first compute $\mathcal{F}_1^{-1}$ vertically and then compute $\mathcal{F}_1^{-1}$ horizontally; these two computations with the reverse order also give the same value.
\QED
\end{example}

\begin{remark}
Because we deduce Propositions \ref{low-dim} and \ref{lower-dim} from Definitions \ref{Generalization} and \ref{inverse-Fourier}, we may conversely define our generalized DFT and IDFT by induction on dimension $N$.
Then, the inductive step is performed using the formulae \eqref{FandF}--\eqref{F-1andF} in Propositions \ref{low-dim} and \ref{lower-dim}.
Moreover, Fourier inversion formulae for general $N$ are deduced from these formulae only for $N=1$, and the equalities \eqref{DFT} and \eqref{inverse} in Definitions \ref{Generalization} and \ref{inverse-Fourier} are also deduced from the simplest cases \eqref{F1} and \eqref{F1i}.
Because Definitions \ref{Generalization} and \ref{inverse-Fourier} provide a procedure to compute each value point by point, one can compute the value on a part of $A$ and $\Omega$ using \eqref{DFT} and \eqref{inverse}.
On the other hand, if we adopt the inductive expressions, then we must compute the value on a whole $A$ or $\Omega$.
These two methods will be compared in Section \ref{Estimation}, and the latter one has less computational complexity in our applications.
\QED
\end{remark}

Another property of our generalized DFT is that its transposed map is equal to the evaluation map of $N$-variable polynomials; see Remark \ref{adjoint}.

%%%%%%%%%%%%%%%%%%%%%%%%%%%%%%%%%%%%%%%%%%%%%%%%%%%%%%%%%%%%%%%%
\section{Main Lemma
\label{Main lemma}}

%%%%%%%%%%%%%%%%%%%%%%%%%%%%%%%%%%%%%%%%%%%%%%%%%%%%%%%%%%%%%%%%
\subsection{Two vector spaces $V_D$ and $V_\Psi$
\label{vector spaces}}

Let $\Psi\subseteq\Omega$ with $\Psi\not=\emptyset$ and $n=|\Psi|$.
One of the two vector spaces in the lemma is given by
\footnote{Because we have used the vector notation $\left(c_{\underline{\omega}}\right)_\Omega\in V_\Omega$ and $V_\Psi$ is a subspace of $V_\Omega$ as noted in Section II, we represent a vector in $V_\Psi$ as $\left(c_{\underline{\psi}}\right)_\Psi$ using the same $c$.}
$$
V_\Psi=\left\{\left.\left(c_{\underline{\psi}}\right)_\Psi\,\right|\,\underline{\psi}\in\Psi,\,c_{\underline{\psi}}\in\mathbb{F}_q\right\},
$$
namely, $V_\Psi$ is the vector space over $\mathbb{F}_q$ indexed by the elements of $\Psi$ whose dimension is trivially $n$.
The other of the two vector spaces is somewhat complicated to define, as it requires Gr\"obner basis theory \cite{Cox97}.
Let $\mathbb{F}_q[\underline{x}]$ be the ring of polynomials with coefficients in $\mathbb{F}_q$ whose variables are $x_1,\cdots,x_N$.
Let $Z_\Psi$ be an ideal of $\mathbb{F}_q[\underline{x}]$ defined by
$$
Z_\Psi=\left\{\left.f(\underline{x})\in\mathbb{F}_q[\underline{x}]\,\right|\,f(\underline{\psi})=0\mbox{ for all }\underline{\psi}\in\Psi\right\}.
$$
Note that $x_i^q-x_i\in Z_\Psi$ for all $1\le i\le N$, as $\Psi\not=\emptyset$.
We fix a monomial order $\preceq$ of $\left\{\left.\underline{x}^{\underline{d}}\,\right|\,\underline{d}\in\mathbb{N}_0^N\right\}$ \cite{Cox97}, and then denote, for $f(\underline{x})\in\mathbb{F}_q[\underline{x}]$,
\begin{align}\label{LM}
&\mathrm{LM}(f)=\max_{\preceq}\left\{\left.\underline{x}^{\underline{d}}\,\right|\,\underline{d}\in\mathbb{N}_0^N,\,f_{\underline{d}}\not=0\right\}
\\
&\mbox{if }\;f(\underline{x})=\sum_{\underline{d}\in\mathbb{N}_0^N,\,f_{\underline{d}}\not=0}f_{\underline{d}}\underline{x}^{\underline{d}}\in\mathbb{F}_q[\underline{x}]\;\mbox{ and }f(\underline{x})\not=0,\nonumber
\end{align}
where $\underline{x}^{\underline{d}}=x_1^{d_1}\cdots x_N^{d_N}$ for $\underline{d}=\left(d_1,\cdots,d_N\right)\in\mathbb{N}_0^N$, and $\mathrm{LM}(f)$ is called the leading monomial of $f(\underline{x})\in\mathbb{F}_q[\underline{x}]$.
The delta set $D=D(\Psi)\subseteq\mathbb{N}_0^N$ of $Z_\Psi$ for $\Psi$ \cite{Saints95} is then defined by
\footnote{The delta set $D=D(\Psi)$ was referred to as a complement of monomial ideals in \cite{Cox97}, and as a footprint in \cite{footprint}.
}
$$
D=D(\Psi)=\left.\mathbb{N}_0^N\right\backslash\left\{\mathrm{mdeg}\left(\mathrm{LM}(f)\right)\,\left|\,0\not=f(\underline{x})\in Z_\Psi\right.\right\},
$$
where $\mathrm{mdeg}\left(\underline{x}^{\underline{d}}\right)=\underline{d}\in\mathbb{N}_0^N$.
Fortunately, $D(\Psi)$ has an intuitive description if a Gr\"obner basis $\mathcal{G}_\Psi$ of $Z_\Psi$ is obtained; it corresponds to the area surrounded by $\mathrm{LM}\left(\mathcal{G}_\Psi\right)$.
The delta set $D=D(\Psi)\subseteq\mathbb{N}_0^N$ of $Z_\Psi$ for $\Psi$ is equivalently defined by
\begin{equation*}
\begin{split}
&\left.\left\{\underline{x}^{\underline{d}}\,\right|\,
\underline{d}\in D(\Psi)\right\}
=\\
&\qquad\left.\left\{\underline{x}^{\underline{d}}\,\left|\,
\underline{d}\in\mathbb{N}_0^N
\right.\right\}
\right\backslash
\left\{\mathrm{LM}(f)\,\left|\,
0\not=f(\underline{x})\in Z_\Psi
\right.\right\}.
\end{split}
\end{equation*}
The other of the two vector spaces is then given by
\footnote{
Because we have used the vector notation $\left(h_{\underline{a}}\right)_A\in V_A$ and $V_D$ is a subspace of $V_A$ as noted in Section II, we represent a vector in $V_D$ as $\left(h_{\underline{d}}\right)_D$ using the same $h$.}
$$
V_D=V_{D(\Psi)}=\left\{\left.\left(h_{\underline{d}}\right)_D\,\right|\,\underline{d}\in D(\Psi),\,h_{\underline{d}}\in\mathbb{F}_q\right\},
$$
namely, the vector space over $\mathbb{F}_q$ indexed by the elements of $D(\Psi)$.
It is known \cite{Fitzgerald-Lax98} that the evaluation map
\begin{equation}\label{evaluation}
\mathrm{ev}:\mathbb{F}_{q}[\underline{x}]/Z_\Psi\to V_\Psi\quad
\left[f\left(\underline{x}\right)\mapsto
\left(f\left(\underline{\psi}\right)\right)_\Psi\right]
\end{equation}
is isomorphic.
\footnote{The proof is quoted from \cite{Fitzgerald-Lax98}; the kernel of ev is trivially $Z_\Psi$ and the image of ev is $V_\Psi$ as, for $\underline{\phi}\in\Psi$, $f_{\underline{\phi}}(\underline{x})=\prod_{i=1}^N\left\{1-\left(x_i-\phi_i\right)^{q-1}\right\}$ satisfies $f_{\underline{\phi}}\left(\underline{\phi}\right)=1$ and $f_{\underline{\phi}}\left(\underline{\psi}\right)=0$ for all $\underline{\psi}\not=\underline{\phi}$.}
Because $\left.\left\{\underline{x}^{\underline{d}}\,\right|\,\underline{d}\in D(\Psi)\right\}$ is a basis of the quotient ring $\mathbb{F}_{q}[\underline{x}]/Z_\Psi$ viewed as a vector space over $\mathbb{F}_{q}$, $\mathbb{F}_{q}[\underline{x}]/Z_\Psi$ is isomorphic to $V_D$.
Thus, the map \eqref{evaluation} can also be written as
\begin{equation}\label{ev}
\mathrm{ev}:V_D\to V_\Psi\quad
\left[\left(h_{\underline{d}}\right)_D\mapsto
\left(\sum_{\underline{d}\in D}h_{\underline{d}}\underline{\psi}^{\underline{d}}\right)_\Psi\right].
\end{equation}
In particular, it follows from the isomorphism \eqref{evaluation} or \eqref{ev} that $\left|D(\Psi)\right|=|\Psi|$ and $\dim_{\mathbb{F}_q}V_D=n$.

Because $V_D$ and $V_\Psi$ have the same dimension $n$, it is trivial that $V_D$ is isomorphic to $V_\Psi$ as a vector space over $\mathbb{F}_q$.
However, this type of isomorphic maps depends on the choice of the bases of the vector spaces; additionally, in coding theory, the normal orthogonal basis is not always convenient for encoding and decoding.
Our lemma asserts that there is a canonical isomorphic map that does not depend on the bases.
As explained in Introduction, the isomorphic map $V_D\to V_\Psi$ of the lemma is given by the composition of the extension defined in the next subsection and the IDFT.

Consider another linear map $\mathcal{P}$ given by
\begin{equation}\label{partial DFT}
\mathcal{P}:V_\Psi\to V_D\quad\left[
\left(c_{\underline{\psi}}\right)_\Psi\mapsto
\left(\sum_{\underline{\psi}\in\Psi}c_{\underline{\psi}}\underline{\psi}^{\underline{d}}\right)_D\right],
\end{equation}
which is called a proper transform \cite{Saints95}.
\begin{proposition}\label{transposed}
{\it(Relation between $\mathrm{ev}$ and $\mathcal{P}$)}
If the normal orthogonal bases are taken as those of $V_D$ and $V_\Psi$, then the two matrices that represent $\mathrm{ev}:V_D\to V_\Psi$ in \eqref{ev} and $\mathcal{P}:V_\Psi\to V_D$ in \eqref{partial DFT} are the transpose of each other.
\QED
\end{proposition}
The proof is described in Appendix \ref{transpose}.
It follows from Proposition \ref{transposed} that $\mathcal{P}$ is also isomorphic; this fact is noted in \cite{Saints95}.

\begin{remark}\label{adjoint}
(Continued from Remark \ref{Remark 2})
In this remark, we consider the case $\Psi=\Omega$, and describe the relation among $\mathrm{ev}$, $\mathcal{P}$, and $\mathcal{F}$.
It follows from $\mathcal{G}_\Omega=\left\{x_i^q-x_i\,|\,1\le i\le N\right\}$ that $D(\Omega)=A$.
Then, $\mathrm{ev}$ becomes the map between $V_A\to V_\Omega$.
On the other hand, $\mathcal{P}:V_\Omega\to V_A$ is equivalent to $\mathcal{F}$.
Thus, Proposition \ref{transposed} implies that $\mathrm{ev}:V_A\to V_\Omega$ is the transposed map of $\mathcal{F}:V_\Omega\to V_A$.
We now demonstrate this fact for the simple case of $N=1$ and $q=8$.
Because $\{1,x,x^2,\cdots,x^7\}$ is a basis of the quotient ring $\mathbb{F}_8[x]/Z_\Omega$ viewed as a vector space over $\mathbb{F}_8$ isomorphic to $V_A$, the matrix representing $\mathrm{ev}:V_A\to V_\Omega$ is equal to $\left[x^{l-1}\left(\omega_m\right)\right]$ with $(l,m)$-th entry $x^{l-1}\left(\omega_m\right)=\left.x^{l-1}\right|
\renewcommand\arraystretch{0.4}
\arraycolsep = 0mm
\begin{array}{l}\\x=\omega_m\end{array}$ for $1\le l,m\le8$, where $\omega_1=0$ and $\omega_m=\alpha^{m-2}$ with $1<m\le8$ are in $\Omega$.
Thus, we can determine $\left(c_{\omega}\right)_\Omega=\mathrm{ev}\left(\left(h_a\right)_A\right)$ by
$$
\left(c_{\omega}\right)_\Omega=\left(h_a\right)_A
\left[
\renewcommand\arraystretch{0.9}
\arraycolsep = 0.8mm
\begin{array}{cccccccc}
1&1&1&1&1&1&1&1\\
0&1&\alpha^1&\alpha^2&\alpha^3&\alpha^4&\alpha^5&\alpha^6\\
0&1&\alpha^2&\alpha^4&\alpha^6&\alpha^1&\alpha^3&\alpha^5\\
0&1&\alpha^3&\alpha^6&\alpha^2&\alpha^5&\alpha^1&\alpha^4\\
0&1&\alpha^4&\alpha^1&\alpha^5&\alpha^2&\alpha^6&\alpha^3\\
0&1&\alpha^5&\alpha^3&\alpha^1&\alpha^6&\alpha^4&\alpha^2\\
0&1&\alpha^6&\alpha^5&\alpha^4&\alpha^3&\alpha^2&\alpha^1\\
0&1&1&1&1&1&1&1
\end{array}
\right].
$$
According to Proposition \ref{transposed}, this matrix is equal to the transpose of the matrix that appeared in Remark \ref{Remark 1}; actually, this can be directly checked.
\QED
\end{remark}

%%%%%%%%%%%%%%%%%%%%%%%%%%%%%%%%%%%%%%%%%%%%%%%%%%%%%%%%%%%%%%%%
\subsection{Extension map $\mathcal{E}:V_D\to V_A$
\label{Extension map}}

Let $\mathcal{G}_\Psi$ be a Gr\"obner basis with respect to $\preceq$ for the ideal $Z_\Psi$.
We assume that $\mathcal{G}_\Psi$ consists of $z$ elements $\{g^{(w)}\}_{0\le w<z}$.
According to Gr\"obner basis theory \cite{Cox97}, we say that the Gr\"obner basis $\mathcal{G}_\Psi$ is reduced if and only if, for all distinct $g_1,g_2\in\mathcal{G}_\Psi$, no monomial appearing in $g_1$ is a multiple of $\mathrm{LM}(g_2)$, and the coefficient of the leading monomial in $g\in\mathcal{G}_\Psi$ is equal to one.
Then, $\mathcal{G}_\Psi=\left\{g^{(w)}\right\}_{0\le w<z}$ becomes of the form
\begin{equation}\label{reduced}
\begin{split}
&g^{(w)}=g^{(w)}(\underline{x})=\\
&\underline{x}^{\underline{a}_w}+\sum_{\underline{d}\in D(\Psi)}g_{\underline{d}}^{(w)}\underline{x}^{\underline{d}}\in Z_\Psi
\;\mbox{ with }\underline{a}_w\in\left.A\right\backslash D(\Psi).
\end{split}
\end{equation}
It is shown \cite{Cox97} that the reduced Gr\"obner basis can be computed from any Gr\"obner basis, and there exists a unique reduced Gr\"obner basis  for each $Z_\Psi$ with respect to a fixed monomial order $\preceq$.
However, we first do not assume that the Gr\"obner basis is reduced, and we deal with an arbitrary Gr\"obner basis for a while.

\begin{definition}\label{map}
{\it(Map from Gr\"obner bases)}
A linear map $\mathcal{E}$ is defined by
\begin{equation}\label{E}
\mathcal{E}:V_D\rightarrow V_A\quad\left[\left(h_{\underline{d}}\right)_D\mapsto\left(h_{\underline{a}}\right)_A\right],
\end{equation}
where, for each $\underline{a}\in A$, $h_{\underline{a}}$ is determined by
\begin{equation}\label{recursion}
h_{\underline{a}}=
\sum_{\underline{d}\in D(\Psi)}
v_{\underline{d}}h_{\underline{d}},
\end{equation}
if the division algorithm by $\mathcal{G}_\Psi$ produces the equality
\begin{equation}\label{reduction}
\underline{x}^{\underline{a}}=\sum_{0\le w<z}u^{(w)}(\underline{x})g^{(w)}(\underline{x})+v(\underline{x})
\end{equation}
for some $u^{(w)}(\underline{x})\in\mathbb{F}_q[\underline{x}]$ for all $0\le w<z$ and for some $v(\underline{x})\in\mathbb{F}_q[\underline{x}]$ with $v(\underline{x})=\sum_{\underline{d}\in D(\Psi)}v_{\underline{d}}\underline{x}^{\underline{d}}$.
\footnote{
For an arbitrary $\underline{a}\in A$, such $u^{(w)}(\underline{x})$ and $v(\underline{x})$ can be always computed through the division algorithm by $\mathcal{G}_\Psi$. Moreover, such $v(\underline{x})$ is uniquely determined for each $\underline{a}\in A$.
For these facts from Gr\"obner basis theory, see, e.g., \cite{Cox97}.}
\QED
\end{definition}
It follows from this definition that, for all $\underline{a}=\underline{d}\in D$, each $h_{\underline{a}}$ of $\left(h_{\underline{a}}\right)_A=\mathcal{E}\left(\left(h_{\underline{d}}\right)_D\right)$ is equal to $h_{\underline{d}}$ of $\left(h_{\underline{d}}\right)_D$, because $v(\underline{x})=\underline{x}^{\underline{d}}$.
This gives the consistency of notation $\left(h_{\underline{a}}\right)_A=\mathcal{E}\left(\left(h_{\underline{d}}\right)_D\right)$ and implies the injectivity of $\mathcal{E}$.

The map $\mathcal{E}$ of Definition \ref{map} enables us to extend syndrome values to DFT, for example, in the decoding of Reed--Solomon (RS) codes as stated below.
If we represent a codeword of a RS code as $c(x)$ and an error polynomial as $e(x)$ as 7.2 of \cite{Blahut83}, we obtain syndrome values $r(\alpha^i)=e(\alpha^i)$ by substituting the roots $\alpha^i$ of the generator polynomial for $r(x)=c(x)+e(x)$.
Then, the syndrome values are equal to a part of DFT $\left(e(\alpha^0),e(\alpha^1),\cdots,e(\alpha^{n-1})\right)$.
The following Proposition \ref{extension} and diagram \eqref{first-diagram} indicate that the whole of DFT is obtained by $\mathcal{E}$ for syndrome values, where more specific description is given at Algorithm \ref{Finding erasures and errors} in \ref{general case}.

\begin{proposition}\label{extension}
{\it(Prolongation via $\mathcal{E}$ for the linear sum of monomial values)}
Let $\left(h_{\underline{d}}\right)_D=\left(\sum_{\underline{\psi}\in\Psi}c_{\underline{\psi}}\underline{\psi}^{\underline{d}}\right)_D\in V_D$ for some $\left(c_{\underline{\psi}}\right)_\Psi\in V_\Psi$ according to the isomorphism $\mathcal{P}:V_\Psi\to V_D$ of \eqref{partial DFT}.
Moreover, let $\left(h_{\underline{a}}\right)_A=\mathcal{E}\left(\left(h_{\underline{d}}\right)_D\right)\in V_A$.
Then, it follows that $\left(h_{\underline{a}}\right)_A=\left(\sum_{\underline{\psi}\in\Psi}c_{\underline{\psi}}\underline{\psi}^{\underline{a}}\right)_A$.
\QED
\end{proposition}
The proof of this proposition is described in Appendix \ref{Proof of Extension}.

We denote by $\mathcal{I}$ the inclusion map
\begin{equation}\label{inclusion}
\mathcal{I}:V_\Psi\rightarrow V_\Omega\quad\left[\left(c_{\underline{\psi}}\right)_\Psi\mapsto\left(c_{\underline{\omega}}\right)_\Omega\right],
\end{equation}
where $c_{\underline{\omega}}=c_{\underline{\psi}}$ if $\underline{\omega}=\underline{\psi}\in\Psi$ and $c_{\underline{\omega}}=0$ if $\underline{\omega}\not\in\Psi$.
Then, Proposition \ref{extension} asserts that the following commutative diagram, i.e., $\mathcal{E}\circ\mathcal{P}=\mathcal{F}\circ\mathcal{I}$, exists.
\def\ext{\mathcal{E}}
\def\inc{\mathcal{I}}
\def\inv{\mathcal{P}}
\def\Ft{\mathcal{F}}
\begin{equation}\label{first-diagram}
\begin{diagram}
V_A & & \lTo^\Ft & & V_{\Omega} \\
\uTo^\ext & &  & & \uTo_\inc \\
V_D & & \lTo^\inv & & V_\Psi \\
\end{diagram}
\end{equation}
\vspace{0.5em}

Furthermore, if we also assume that the Gr\"obner basis $\mathcal{G}_\Psi$ is reduced, then we obtain an alternative description of the extension map $\mathcal{E}:V_D\rightarrow V_A$.
From now on, $A=\{0,1,\cdots,q-1\}^N$ is considered as a semigroup by the component-wise addition $\underline{a}+\underline{b}$ for $\underline{a}=\left(a_1,\cdots,a_N\right),\underline{b}=\left(b_1,\cdots,b_N\right)\in A$, where the component $a_i+b_i$ is viewed within $1\le(a_i+b_i\,\mathrm{mod}\,(q-1))<q$ if $a_i+b_i\not=0$ for $1\le i\le N$.
For example, $(0,0,1,2)+(0,3,1,2)=(0,3,2,1)$ in $A$ if $N=4$ and $q=4$.
This semigroup structure of $A$ comes naturally from the multiplication of monomials in $\mathbb{F}_{q}[\underline{x}]/Z_\Omega$, which is isomorphic to $V_A$ as a vector space because $D(\Omega)=A$.
Moreover, for $\underline{a},\underline{b}\in A$, we denote $\underline{a}\ge\underline{b}$ if $a_i\ge b_i$ component-wise for all $1\le i\le N$, or equivalently, if there is $\underline{c}\in A$ such that $\underline{a}=\underline{b}+\underline{c}$.

\begin{proposition}\label{recurrence}
{\it(M-D linear feedback shift registers from Gr\"obner bases)}
Suppose that the Gr\"obner basis $\mathcal{G}_\Psi=\left\{g^{(w)}\right\}_{0\le w<z}$ is reduced and of the form \eqref{reduced}.
If $\left(h_{\underline{a}}\right)_A=\mathcal{E}\left(\left(h_{\underline{d}}\right)_D\right)$ for some $\left(h_{\underline{d}}\right)_D\in V_D$ with $D=D(\Psi)$, then we have, for all $\underline{a}\in A$ and all $0\le w<z$,
\begin{equation}\label{generation}
h_{\underline{a}}=\left\{
\begin{array}{ll}
h_{\underline{d}} & \underline{a}=\underline{d}\in D\\
-\sum_{\underline{d}\in D}
g_{\underline{d}}^{(w)}h_{\underline{a}+\underline{d}-\underline{a}_w} & \underline{a}\ge\underline{a}_w.
\end{array}\right.
\end{equation}
Conversely, if $\left(h_{\underline{a}}\right)_A\in V_A$ satisfies that, for each $\underline{a}\in A\backslash D(\Psi)$, there exists at least one $0\le w<z$ such that \eqref{generation}, then we have $\left(h_{\underline{a}}\right)_A=\mathcal{E}\left(\left(h_{\underline{d}}\right)_D\right)$ for $\left(h_{\underline{d}}\right)_D\in V_D$ with $D=D(\Psi)$ and \eqref{generation}.
\QED
\end{proposition}
The proof is described in Appendix \ref{Proof of recurrence}.
To actually compute the value of $\mathcal{E}\left(\left(h_{\underline{d}}\right)_D\right)=\left(h_{\underline{a}}\right)_A$ from a given $\left(h_{\underline{d}}\right)_D$, we can generate $(h_{\underline{a}})_A$ inductively by \eqref{generation}, because, for each $\underline{a}\in A\backslash D(\Psi)$, at least one $0\le w<z$ can be chosen such that $\underline{a}\ge\underline{a}_w$.
Moreover, the induction to generate $(h_{\underline{a}})_A$ works because the monomial order is a total order \cite{Cox97} and we have $\underline{a}\succeq\underline{a}+\underline{d}-\underline{a}_w$ and $\underline{a}\not=\underline{a}+\underline{d}-\underline{a}_w$ in case of $\underline{a}\ge\underline{a}_w$ in the right-hand side of \eqref{generation}.
Then, the latter half of Proposition \ref{recurrence} asserts that the resulting value does not depend on the choice and order of the generation, and that $\mathcal{E}\left(\left(h_{\underline{d}}\right)_D\right)=\left(h_{\underline{a}}\right)_A$ is uniquely determined.
In the rest of the paper, for simplicity, we adopt \eqref{generation} to compute the value of $\mathcal{E}\left(\left(h_{\underline{d}}\right)_D\right)=\left(h_{\underline{a}}\right)_A$ in place of \eqref{recursion} and \eqref{reduction}.

%%%%%%%%%%%%%%%%%%%%%%%%%%%%%%%%%%%%%%%%%%%%%%%%%%%%%%%%%%%%%%%%
\subsection{Isomorphic map $\mathcal{C}:V_D\to V_\Psi$
\label{Isomorphic map}}

From now on, we denote $\mathcal{R}$ as the restriction map
\begin{equation}\label{restriction}
\mathcal{R}:V_\Omega\rightarrow V_\Psi\quad\left[\left(c_{\underline{\omega}}\right)_{\Omega}\mapsto\left(c_{\underline{\psi}}\right)_{\Psi}\right].
\end{equation}
It follows from \eqref{first-diagram} that $\mathcal{F}^{-1}\circ\mathcal{E}\circ\mathcal{P}=\mathcal{I}$.
Moreover, $\mathcal{R}\circ\mathcal{F}^{-1}\circ\mathcal{E}\circ\mathcal{P}=\mathcal{R}\circ\mathcal{I}$ is the identity map on $V_\Psi$.
This leads to the following lemma, which is frequently used in this paper.
\begin{Main}
Let $\mathcal{G}_\Psi$ be a Gr\"obner basis of $Z_\Psi$ for $\Psi\subseteq\Omega$, and let $\mathcal{E}:V_D\to V_A$ be the extension map defined by \eqref{E}.
Then, the composition map $\mathcal{C}=\mathcal{R}\circ\mathcal{F}^{-1}\circ\mathcal{E}:V_D\to V_\Psi$ in the following commutative diagram gives an isomorphism between $V_D$ and $V_\Psi$.
\def\rest{\mathcal{R}}
\def\comp{\mathcal{C}}
\def\Fti{\mathcal{F}^{-1}}
\begin{diagram}
V_A & & \rTo^\Fti & & V_{\Omega} \\
\uTo^\ext & &  & & \dTo_\rest \\
V_D & & \rTo^\comp & & V_\Psi \\
\end{diagram}
Moreover, we have that
\begin{equation}\label{vanish}
\left(c_{\underline{\omega}}\right)_\Omega\in\mathcal{F}^{-1}\left(\mathcal{E}\left(V_D\right)\right)\:\Longrightarrow\:c_{\underline{\omega}}=0\,\mbox{ for all }\,\underline{\omega}\in\Omega\backslash\Psi.\;\;\QED
\end{equation}
\end{Main}

\begin{remark}
As $\mathcal{C}=\mathcal{P}^{-1}$, our $\mathcal{C}$ can also be obtained from the multiplication of the inverse matrix representing \eqref{partial DFT}.
However, if $\Psi$ is changed, then the inverse matrix must be computed each time.
As $\Psi$ takes, e.g., the set of erasure-and-error locations and $\mathcal{C}$ has a lower computational complexity order than Gaussian elimination, there are many cases where $\mathcal{C}$ outperforms computing the inverse matrix, as shown in Section \ref{Estimation}.
\QED
\end{remark}

\begin{remark}
The above proof of our Main Lemma can also be applied to the non-zero indexed case \cite{ISIT07},\cite{ISITA12} where $A=\{0,1,\cdots,q-2\}^N$ has a cyclic structure mod $(q-1)$ and $\Omega=\left(\mathbb{F}_q^\times\right)^N$.
\QED
\end{remark}

\begin{example}\label{RS}
Putting $N=1$, $q=8$, and $\alpha\in\mathbb{F}_8$ with $\alpha^3+\alpha+1=0$, consider the natural order $\preceq$ to be a monomial order, i.e., $0\preceq1\preceq2\preceq\cdots\preceq7$ on $A=\{0,1,\cdots,7\}$.
Choose $\Psi\subseteq\Omega=\{0,1,\alpha,\alpha^2,\cdots,\alpha^6\}$ as $\Psi=\{0,\alpha,\alpha^3,\alpha^6\}$.
Then, $D=D(\Psi)=\{0,1,2,3\}$ and $\mathcal{G}_\Psi=\{g(x)\}$, where
$$
g(x)=\prod_{\psi\in\Psi}(x-\psi)=\alpha^3x+\alpha^3x^2+\alpha^2x^3+x^4.
$$
For $\left(h_d\right)_D=\left(h_0,h_1,h_2,h_3\right)=(\alpha^2,\alpha^3,\alpha^5,\alpha^0)$, $\mathcal{E}\left(\left(h_d\right)_D\right)=\left(h_a\right)_A=\left(h_0,h_1,\cdots,h_7\right)$ is given by
$$
\left(h_0,h_1,\cdots,h_7\right)=\left(\alpha^2,\alpha^3,\alpha^5,\alpha^0,\alpha^3,\alpha^4,\alpha^3,\alpha^3\right),
$$
where, e.g., $h_4=\alpha^3\alpha^3+\alpha^3\alpha^5+\alpha^2\alpha^0=\alpha^3$, and $\mathcal{F}^{-1}\left(\mathcal{E}\left(\left(h_d\right)_D\right)\right)=\left(c_{\omega}\right)_\Omega$ is given by
$$
\left(c_{0},c_{1},c_{\alpha},\cdots,c_{\alpha^6}\right)=
\left(\alpha^5,0,\alpha^2,0,1,0,0,\alpha^4\right).
$$
Note that $c_{\omega}=0$ if $\omega\not\in\Psi$.
Then, $\mathcal{C}\left(\left(h_d\right)_D\right)=\left(c_{\psi}\right)_\Psi=\left(c_{0},c_{\alpha},c_{\alpha^3},c_{\alpha^6}\right)=\left(\alpha^5,\alpha^2,1,\alpha^4\right)$.
\QED
\end{example}

\begin{example}\label{example-cross}
\begin{figure}[t!]% Fig. 2
\centering
  \resizebox{7.5cm}{!}{\includegraphics{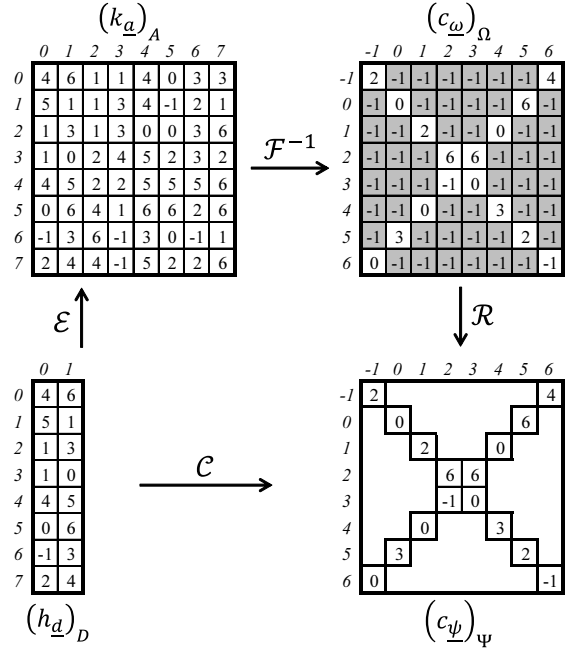}}
\caption{Numerical example of Main Lemma, where $\Psi$ is given by \eqref{itimatu} in Example \ref{example-cross}.
The value of $\left(c_{\underline{\omega}}\right)_{\Omega}$ in the shaded box indicates $c_{\underline{\omega}}$ on $\Omega$ outside the $\Psi$ of \eqref{itimatu}. Note that these values are all $-1$ according to assertion \eqref{vanish} of Main Lemma.
\label{figure-cross}}
\end{figure}
Putting $N=2$, $q=8$, and $\alpha\in\mathbb{F}_8$ with $\alpha^3+\alpha+1=0$, consider the lexicographic order $\preceq$ to be a monomial order, i.e., $(0,0)\preceq(1,0)\preceq(2,0)\preceq\cdots\preceq(7,0)\preceq(0,1)\preceq(1,1)\preceq\cdots\preceq(7,7)$ on $A=\{0,1,\cdots,7\}^2$.
Choose $\Psi\subseteq\Omega=\mathbb{F}_8^2$ as
\begin{equation}\label{itimatu}
\Psi=\left\{
\begin{array}{l}
(0,0),(0,\alpha^6),
(\alpha^0,\alpha^0),(\alpha^0,\alpha^5),\\
(\alpha^1,\alpha^1),(\alpha^1,\alpha^4),
(\alpha^2,\alpha^2),(\alpha^2,\alpha^3),\\
(\alpha^3,\alpha^2),(\alpha^3,\alpha^3),
(\alpha^4,\alpha^1),(\alpha^4,\alpha^4),\\
(\alpha^5,\alpha^0),(\alpha^5,\alpha^5),
(\alpha^6,0),(\alpha^6,\alpha^6)
\end{array}
\right\},
\end{equation}
which, in order to show a pictorial example, is the cross pattern $\left(c_{\underline{\omega}}\right)_\Omega$ in Fig.\ \ref{figure-cross}.
We denote $x=x_1$ and $y=x_2$ in $\mathbb{F}_8[\underline{x}]=\mathbb{F}_8[x_1,x_2]$.
An element $g(x,y)\in\mathcal{G}_\Psi$ of the Gr\"obner basis can then be characterized as $g(x,y)\in\mathbb{F}_8[x,y]$ with $g(\omega_1,\omega_2)=0$ for all $(\omega_1,\omega_2)\in\Psi$ that has the minimum $\mathrm{LM}(g)$ with respect to $\preceq$.
One of $\mathcal{G}_{\Psi}$ is computed as
\begin{align*}
&g(x,y)=\alpha^6x+\alpha^0x^2+\alpha^1x^3+\alpha^2x^4+\alpha^3x^5+\alpha^4x^6\\
&\quad+y\left(\alpha^6+\alpha^1x^2+\alpha^2x^3+\alpha^3x^4+\alpha^4x^5\right)+y^2.
\end{align*}
The other elements of $\mathcal{G}_{\Psi}$ are not necessary to extend $\left(h_{\underline{d}}\right)_D$ because of the semigroup structure of $A$.
For a given $\left(h_{\underline{d}}\right)_D$, all values of Main Lemma are shown in Fig.\ \ref{figure-cross}.
For example, $h_{(2,2)}$ is generated as
$$
h_{(2,2)}=-\sum_{\underline{d}\in D(\Psi)}
g_{\underline{d}}h_{(2,2)+\underline{d}-(0,2)},
$$
where it should be noted that $h_{(2,2)+(6,0)-(0,2)}=h_{(1,0)}$. Thus, we have
\begin{align*}
&h_{(2,2)}=\alpha^6\alpha^1+\alpha^0\alpha^4+\alpha^1\alpha^0+\alpha^20+\alpha^3\alpha^2+\alpha^4\alpha^5\\
&\quad+\left(\alpha^6\alpha^3+\alpha^1\alpha^5+\alpha^2\alpha^6+\alpha^3\alpha^3+\alpha^4\alpha^4\right)=\alpha^1.\quad\QED
\end{align*}
The data of $\mathcal{F}^{-1}\left(\left(h_{\underline{a}}\right)_A\right)=\left(c_{\underline{\omega}}\right)_\Omega$ has already been treated in Example \ref{multi} and Fig.\ \ref{multidimensional}.
\end{example}

%%%%%%%%%%%%%%%%%%%%%%%%%%%%%%%%%%%%%%%%%%%%%%%%%%%%%%%%%%%%%%%%
\section{Applications of Main Lemma
\label{Application}}

%%%%%%%%%%%%%%%%%%%%%%%%%%%%%%%%%%%%%%%%%%%%%%%%%%%%%%%%%%%%%%%%
\subsection{Affine variety codes \cite{Fitzgerald-Lax98}
\label{arbitrary subset}}

Let $\Psi\subseteq\Omega$ with $\Psi\not=\emptyset$ and $n=|\Psi|$, as at the beginning of Subsection \ref{vector spaces}.
Let $U$ be a subspace of $V_{D(\Psi)}$.
Consider an affine variety code \cite{Fitzgerald-Lax98} with code length $n$
\begin{align}\label{ev image}
C(U,\Psi)&=\mathrm{ev}(U)\\\nonumber
&=\left\{\left(c_{\underline{\psi}}\right)_\Psi\in V_\Psi\left|
\begin{array}{l}\displaystyle
\left(\sum_{\underline{d}\in D}h_{\underline{d}}\underline{\psi}^{\underline{d}}\right)_\Psi=\left(c_{\underline{\psi}}\right)_\Psi\\
\mbox{for some }\left(h_{\underline{d}}\right)_D\in U
\end{array}
\right.\right\},
\end{align}
where $\underline{\psi}^{\underline{d}}=\psi_1^{d_1}\cdots\psi_N^{d_N}$ is as in \eqref{DFT}.
Moreover, consider a dual affine variety code \cite{Fitzgerald-Lax98} with code length $n$
\begin{align}\label{dual code}
C^\perp(U,\Psi)&=\mathrm{ev}(U)^\perp\\
&=\left\{\left(c_{\underline{\psi}}\right)_\Psi\in V_\Psi\left|
\begin{array}{l}\displaystyle
\sum_{\underline{\psi}\in\Psi}c_{\underline{\psi}}\sum_{\underline{d}\in D}h_{\underline{d}}\underline{\psi}^{\underline{d}}=0\\
\mbox{for all }\left(h_{\underline{d}}\right)_D\in U
\end{array}
\right.\right\},\nonumber
\end{align}
where $\sum_{\underline{\psi}\in\Psi}c_{\underline{\psi}}\sum_{\underline{d}\in D}h_{\underline{d}}\underline{\psi}^{\underline{d}}$ in \eqref{dual code} is equal to the inner product of $\left(c_{\underline{\psi}}\right)_\Psi$ and $\mathrm{ev}\left(\left(h_{\underline{d}}\right)_D\right)=\left(\sum_{\underline{d}\in D}h_{\underline{d}}\underline{\psi}^{\underline{d}}\right)_\Psi$ in $V_\Psi$.
Thus, the dimension or number of information symbols $k$ of $C^\perp(U,\Psi)$ is equal to $n-\dim_{\mathbb{F}_q}U$; in other words, $n-k=\dim_{\mathbb{F}_q}U$.
Note that, as vector spaces, these code definitions do not depend on the choice of monomial order; $U\subseteq V_{D(\Psi)}$ is equivalent to $U\subseteq\mathbb{F}_{q}[\underline{x}]/Z_\Psi$.

On the other hand, let $U^\perp$ be the orthogonal complement of $U$ in $V_D$, i.e.,
$$
U^\perp=\left\{\left(h_{\underline{d}}\right)_D\in V_D\left|\,\sum_{\underline{d}\in D}h_{\underline{d}}h'_{\underline{d}}=0\mbox{ for all }\left(h'_{\underline{d}}\right)_D\in U\right\}\right..
$$
Then, similarly to \eqref{ev image}, we obtain
\begin{equation}\label{C image}
C^\perp(U,\Psi)=\mathcal{C}\left(U^\perp\right),
\end{equation}
a proof of which is given in Appendix \ref{Proof of orthogonal complement}.
Whereas the definition \eqref{dual code} of $C^\perp(U,\Psi)$ is indirect and not constructive, the equality \eqref{C image} provides a direct construction.
Moreover, the equality \eqref{C image} corresponds to the non-systematic encoding of $C^\perp(U,\Psi)$.
Actually, non-systematic encoding is obtained, for all $\left(h_{\underline{d}}\right)_D\in U^\perp$, by $\left(c_{\underline{\psi}}\right)_\Psi=\mathcal{C}\left(\left(h_{\underline{d}}\right)_D\right)\in C^\perp(U,\Psi)$ as \eqref{C image}.

\begin{example}
(Continued from Example \ref{RS})
Let $U\subseteq V_D$ be a vector space generated by $\left(1,0,\alpha^4,\alpha^5\right)$ and $\left(0,1,0,\alpha^6\right)$.
If these are represented as polynomials $f(x)=1+\alpha^4x^2+\alpha^5x^3$ and $x+\alpha^6x^3$, then $\mathrm{ev}(U)\subseteq V_\Psi$ is generated by
\begin{align*}
&\left(f(0),f(\alpha),f(\alpha^3),f(\alpha^6)\right)\\
&=\left(1,\alpha^4,\alpha^3,1\right)\mbox{ and }
\left(0,\alpha^4,1,\alpha^4\right).
\end{align*}
Then, $U^\perp\subseteq V_D$ is equal to a vector space generated by
$$
\left(h_0,h_1,h_2,h_3\right)=\left(\alpha^4,0,1,0\right)\mbox{ and }
\left(\alpha^5,\alpha^6,0,1\right).
$$
These extensions are equal to
$$
\left(h_4,h_5,h_6,h_7\right)=\left(\alpha^3,\alpha^2,\alpha^3,\alpha^6\right)\mbox{ and }
\left(0,\alpha^3,\alpha^2,\alpha^3\right).
$$
Thus, $\mathcal{C}\left(U^\perp\right)$ is generated by
$$
\left(c_{0},c_{\alpha},c_{\alpha^3},c_{\alpha^6}\right)=\left(\alpha^3,\alpha^5,\alpha^5,\alpha^6\right)\mbox{ and }
\left(\alpha^2,\alpha^4,\alpha^2,1\right).
$$
The orthogonality is valid, e.g., $\alpha^3+\alpha^2+\alpha+\alpha^6=0$.
\QED
\end{example}

\begin{remark}\label{typical}
A typical case of $U$ is $U=V_B$ for some $B\subseteq D(\Psi)$.
Then, $U^\perp=V_{D\backslash B}$, where $V_B$ and $V_{D\backslash B}$ are considered subspaces of $V_D$, as in Section \ref{Notation}.
\QED
\end{remark}

\begin{example}\label{example-non-systematic-encoding-Hermitian}
\begin{figure}[t!]% Fig. 3
\centering
  \resizebox{7.5cm}{!}{\includegraphics{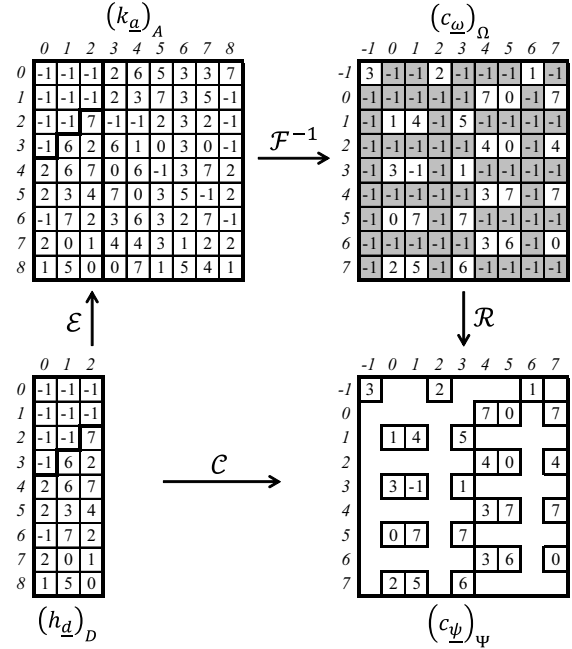}}
\caption{Numerical example of the non-systematic encoding of a Hermitian code over $\mathbb{F}_9$ in Example \ref{example-non-systematic-encoding-Hermitian}. The non-zero elements of $\mathbb{F}_9$ are represented by the number of powers of a primitive element $\alpha$ with $\alpha^2+\alpha=1$, i.e., $0,1,\cdots,7$ means $\alpha^0,\alpha^1,\cdots,\alpha^7$, respectively, and $-1$ means $0\in\mathbb{F}_9$. The value of $\left(c_{\underline{\omega}}\right)_{\Omega}$ not in the shaded box indicates $c_{\underline{\omega}}$ on the $\Psi$ given by \eqref{rational}, and $\left(c_{\underline{\psi}}\right)_{\Psi}$ is a codeword of a Hermitian code.
\label{figure-non-systematic-encoding-Hermitian}}
\end{figure}
Throughout the rest of Section \ref{Application}, we consider a Hermitian code, i.e., a code on the $\mathbb{F}_9$-rational points of a Hermitian curve, in order to compare our method with conventional methods for algebraic geometry codes.
Putting $N=2$, $q=9$, and $\alpha\in\mathbb{F}_9$ with $\alpha^2+\alpha-1=0$, consider the weighted graded lexicographic order \cite{Cox97} to be a monomial order $\preceq$ such that $(a_1,a_2)\preceq(a_1',a_2')\Leftrightarrow 3a_1+4a_2<3a_1'+4a_2'\mbox{ or }3a_1+4a_2=3a_1'+4a_2',\,a_2\le a_2'$, i.e., $(0,0)\preceq(1,0)\preceq(0,1)\preceq(2,0)\preceq(1,1)\preceq(0,2)\preceq\cdots\preceq(1,2)\preceq(4,0)\preceq(0,3)\preceq(3,1)\preceq\cdots\preceq(8,8)$ on $A=\{0,1,\cdots,8\}^2$.
Choose $\Psi\subseteq\Omega=\mathbb{F}_9^2$ as
\begin{equation}\label{rational}
\Psi=\left\{
\begin{array}{l}
(0,0),(0,\alpha^2),(0,\alpha^6),(\alpha^0,\alpha^4),\\
(\alpha^0,\alpha^5),(\alpha^0,\alpha^7),(\alpha^1,\alpha^0),(\alpha^1,\alpha^1),\\
(\alpha^1,\alpha^3),(\alpha^2,\alpha^4),(\alpha^2,\alpha^5),(\alpha^2,\alpha^7),\\
(\alpha^3,\alpha^0),(\alpha^3,\alpha^1),(\alpha^3,\alpha^3),(\alpha^4,\alpha^4),\\
(\alpha^4,\alpha^5),(\alpha^4,\alpha^7),(\alpha^5,\alpha^0),(\alpha^5,\alpha^1),\\
(\alpha^5,\alpha^3),(\alpha^6,\alpha^4),(\alpha^6,\alpha^5),(\alpha^6,\alpha^7),\\
(\alpha^7,\alpha^0),(\alpha^7,\alpha^1),(\alpha^7,\alpha^3),
\end{array}
\right\},
\end{equation}
which agrees with $\left\{\left.(\omega_1,\omega_2)\in\Omega\,\right|\omega_1^4=\omega_2^3+\omega_2\right\}$, a set of $\mathbb{F}_9$-rational points of a Hermitian curve with defining equation $x^4=y^3+y$, where we denote $x=x_1$ and $y=x_2$.
In this case, one of the elements in the Gr\"obner basis $\mathcal{G}_{\Psi}$ is equal to $g(x,y)=y^3-x^4+y$ and the delta set $D(\Psi)$ of $\mathcal{G}_{\Psi}$ is $\left\{\left.\left(a_1,a_2\right)\in A\right|a_2\le2\right\}$.
The other elements of $\mathcal{G}_{\Psi}$ are not necessary to extend $\left(h_{\underline{d}}\right)_D$ because of the semigroup structure of $A$.
Let $B\subseteq D(\Psi)$ be $B=\left\{\left.\left(b_1,b_2\right)\in D(\Psi)\right|3b_1+4b_2\le11\right\}$ and let $U=V_B$.
Then, $C^\perp(U,\Psi)=\mathcal{C}\left(V_{D(\Psi)\backslash B}\right)$ agrees with $C_\Omega(D,mP_\infty)=C_L(D,mP_\infty)^\perp$ in the usual notation \cite{Stichtenoth} for $m=11$ and $D=\sum_{(\omega_1,\omega_2)\in\Psi}P_{\omega_1,\omega_2}$ with $P_{\omega_1,\omega_2}=(\omega_1,\omega_2)$.
For a given $\left(h_{\underline{d}}\right)_D$, all values of Main Lemma are shown in Fig.\ \ref{figure-non-systematic-encoding-Hermitian}, where the vertical axis and the horizontal axis ({\it 0,1,$\cdots$,8}) in $\left(h_{\underline{a}}\right)_A$ indicate $a_1$ and $a_2$ of $\underline{a}=(a_1,a_2)\in A$, and those axes ({\it -1,0,$\cdots$,7}) in $\left(c_{\underline{\omega}}\right)_{\Omega}$ indicate $\omega_1$ and $\omega_2$ of $\underline{\omega}=(\omega_1,\omega_2)\in\Omega$.
\QED
\begin{figure}[t!]% Fig. 4
\centering
  \resizebox{7.5cm}{!}{\includegraphics{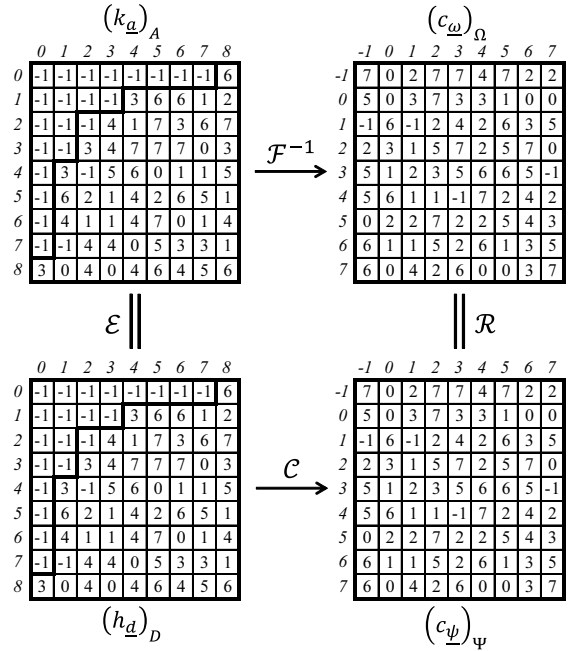}}
\caption{Numerical example for the non-systematic encoding of an extended HCRS code over $\mathbb{F}_9$ in Example \ref{example-non-systematic-encoding-HCRS}. The elements of $\mathbb{F}_9$ are represented as in Fig.\ \ref{figure-non-systematic-encoding-Hermitian}. Because $A=D$ and $\Omega=\Psi$ in this case, $\mathcal{E}$ and $\mathcal{R}$ are identity maps. $\left(c_{\underline{\psi}}\right)_{\Psi}$ is a codeword of an extended HCRS code.
\label{figure-non-systematic-encoding-HCRS}}
\end{figure}
\end{example}
\begin{example}\label{example-non-systematic-encoding-HCRS}
Throughout the rest of Section \ref{Application}, we consider an extended hyperbolic cascaded Reed--Solomon (HCRS) code, which is an example of affine variety codes that are not algebraic geometry codes.
Putting $N=2$, $q=9$, and $\alpha\in\mathbb{F}_9$ with $\alpha^2+\alpha-1=0$, choose $\Psi=\Omega=\mathbb{F}_9^2$ and $A=\{0,1,\cdots,8\}^2$; then, $D=D(\Psi)=A$.
Let $B\subseteq D(\Psi)$ be $B=\left\{\left.\left(b_1,b_2\right)\in A\right|(b_1+1)(b_2+1)<9\right\}$, and let $U=V_B$.
Then, $C^\perp(U,\Omega)=\mathcal{C}\left(V_{A\backslash B}\right)$ is an extended HCRS code \cite{Geil01},\cite{Hoholdt98},\cite{Saints93},\cite{Saints95}.
For a given $\left(h_{\underline{d}}\right)_D$, all values of Main Lemma are shown in Fig.\ \ref{figure-non-systematic-encoding-HCRS}.
\QED
\end{example}

In Subsection \ref{Systematic encoding}, it is shown that Main Lemma also gives the systematic encoding of a class of dual affine variety codes.

%%%%%%%%%%%%%%%%%%%%%%%%%%%%%%%%%%%%%%%%%%%%%%%%%%%%%%%%%%%%%%%%
\subsection{Erasure-and-error decoding: non-systematic case
\label{non-systematic case}}

Henceforth, consider the situation $U=V_B$ with some $B\subseteq D(\Psi)$ from Remark \ref{typical}.
In this subsection, suppose that $\left(h_{\underline{d}}\right)_D\in V_{D\backslash R}$ is encoded into $\left(c_{\underline{\psi}}\right)_\Psi=\mathcal{C}\left(\left(h_{\underline{d}}\right)_D\right)\in C^\perp(V_B,\Psi)$, and consider the decoding problem for this non-systematic encoding.

Suppose also that erasure-and-error $\left(e_{\underline{\psi}}\right)_\Psi\in V_\Psi$ has occurred in a received word $\left(r_{\underline{\psi}}\right)_\Psi=\left(c_{\underline{\psi}}\right)_\Psi+\left(e_{\underline{\psi}}\right)_\Psi\in V_\Psi$ from some channel.
Let $\Phi_1\subseteq\Psi$ be the set of erasure locations and $\Phi_2\subseteq\Psi$ be the set of error locations with $\Phi_1\cap\Phi_2=\emptyset$; we suppose that $\Phi_1$ is known, but $\Phi_2$ and $\left(e_{\underline{\psi}}\right)_\Psi$ are unknown, that $e_{\underline{\psi}}\not=0\Rightarrow\underline{\psi}\in\Phi_1\cup\Phi_2$, and that $\underline{\psi}\in\Phi_2\Rightarrow e_{\underline{\psi}}\not=0$.
We might permit $e_{\underline{\psi}}=0$ for some $\underline{\psi}\in\Phi_1$.
If $|\Phi_1|+2|\Phi_2|<d_\mathrm{FR}$ is valid, where $d_\mathrm{FR}$ denotes the Feng--Rao minimum distance bound \cite{Andersen-Geil08},\cite{BMS05},\cite{Matsumoto-Miura2000},\cite{Salazar-Dunn-Graham}, then it is known that the erasure-and-error version \cite{Koetter98},\cite{Sakata-erasure98} of the BMS algorithm \cite{generic06},\cite{BMS05} or the multidimensional Berlekamp--Massey algorithm calculates the Gr\"obner basis $\mathcal{G}_{\Phi_1\cup\Phi_2}$.
The main difference between the erasure-and-error and ordinary error-only algorithms is in the initialization; as $\Phi_1$ is known, $\mathcal{G}_{\Phi_1}$ can be calculated in advance by the ordinary error-only version, and then $\mathcal{G}_{\Phi_1\cup\Phi_2}$ can be calculated by the erasure-and-error version from the syndrome and the initial value $\mathcal{G}_{\Phi_1}$.
Using the recurrence from $\mathcal{G}_{\Phi_1\cup\Phi_2}$ and Main Lemma, the erasure-and-error decoding algorithm is realized as follows.
\footnote{
In the following algorithms, we use the auxiliary vector notation $\left(v_{\underline{b}}\right)_B\in V_B$, $\left(\widetilde{r}_{\underline{d}}\right)_D\in V_D$, and $\left(k_{\underline{d}}\right)_D\in V_D$.}

\begin{algorithm}{\it (Decoding of non-systematic codewords)}
\label{Decoding of non-systematic codewords}
\begin{description}
 \setlength{\itemsep}{1mm}
\item[Input:]\ $\Phi_1$ and a received word $\left(r_{\underline{\psi}}\right)_\Psi\in V_\Psi$
\item[Output:]\ \ $\left(h_{\underline{d}}\right)_D\in V_{D\backslash B}$ such that $\mathcal{C}\left(\left(h_{\underline{d}}\right)_D\right)=\left(c_{\underline{\psi}}\right)_\Psi\in V_\Psi$
\item[Step 1.]\ $\left(v_{\underline{b}}\right)_B=\left(\sum_{\underline{\phi}\in\Phi_1}\underline{\phi}^{\underline{b}}\right)_B\in V_B$
\item[Step 2.]\ Calculate $\mathcal{G}_{\Phi_1}$ from syndrome $\left(v_{\underline{b}}\right)_B$
\item[Step 3.]\ $\left(\widetilde{r}_{\underline{d}}\right)_D=\left(\sum_{\underline{\psi}\in\Psi}r_{\underline{\psi}}\underline{\psi}^{\underline{d}}\right)_D\in V_D$
\item[Step 4.]\ Calculate $\mathcal{G}_{\Phi_1\cup\Phi_2}$ from $\left(\widetilde{r}_{\underline{b}}\right)_B\in V_B$ and $\mathcal{G}_{\Phi_1}$
\item[Step 5.]\ $\left(k_{\underline{d}}\right)_D=\mathcal{E}\left(\left(\widetilde{r}_{\underline{b}}\right)_B\right)\in V_D$ by $\mathcal{G}_{\Phi_1\cup\Phi_2}$
\item[Step 6.]\ $\left(h_{\underline{d}}\right)_D=\left(\widetilde{r}_{\underline{d}}\right)_D-\left(k_{\underline{d}}\right)_D\in V_{D\backslash B}$
\quad\QED
\end{description}
\end{algorithm}
\vspace{0.5em}
At Step 5, $\left(k_{\underline{d}}\right)_D=\mathcal{E}\left(\left(\widetilde{r}_{\underline{b}}\right)_B\right)$ means that $\left(k_{\underline{d}}\right)_{D(\Psi)}=\mathcal{E}\left(\left(\widetilde{r}_{\underline{d}}\right)_{D(\Phi_1\cup\Phi_2)}\right)$, where the values of $\mathcal{E}$ are only computed on $D(\Psi)\subseteq A$ by the recurrence relation \eqref{generation}.

The validity of this algorithm is proved by the following argument.
It follows from Main Lemma that $\mathcal{C}\left(\left(h_{\underline{d}}\right)_D\right)=\left(c_{\underline{\psi}}\right)_\Psi\Longleftrightarrow\left(h_{\underline{d}}\right)_D=\mathcal{C}^{-1}\left(\left(c_{\underline{\psi}}\right)_\Psi\right)=\left(\sum_{\underline{\psi}\in\Psi}c_{\underline{\psi}}\underline{\psi}^{\underline{d}}\right)_D$.
As $\left(r_{\underline{\psi}}\right)_\Psi=\left(c_{\underline{\psi}}\right)_\Psi+\left(e_{\underline{\psi}}\right)_\Psi$, we have $\left(\widetilde{r}_{\underline{d}}\right)_D=\left(h_{\underline{d}}\right)_D+\left(\sum_{\underline{\psi}\in\Psi}e_{\underline{\psi}}\underline{\psi}^{\underline{d}}\right)_D$ in Step 3 and $\left(\widetilde{r}_{\underline{b}}\right)_B=\left(\sum_{\underline{\psi}\in\Psi}e_{\underline{\psi}}\underline{\psi}^{\underline{b}}\right)_B$ by \eqref{dual code}.
It follows from the proof of Proposition \ref{extension} that $\mathcal{E}\left(\left(\widetilde{r}_{\underline{b}}\right)_B\right)=\left(\sum_{\underline{\psi}\in\Psi}e_{\underline{\psi}}\underline{\psi}^{\underline{d}}\right)_D$ in Step 5, because $U=V_B$ is assumed.
Thus, we obtain $\left(h_{\underline{d}}\right)_D=\left(\widetilde{r}_{\underline{d}}\right)_D-\left(k_{\underline{d}}\right)_D$ in Step 6.

\begin{example}\label{example-non-systematic-decoding-Hermitian}
\begin{figure}[t!]% Fig. 5
\centering
  \resizebox{16cm}{!}{\includegraphics{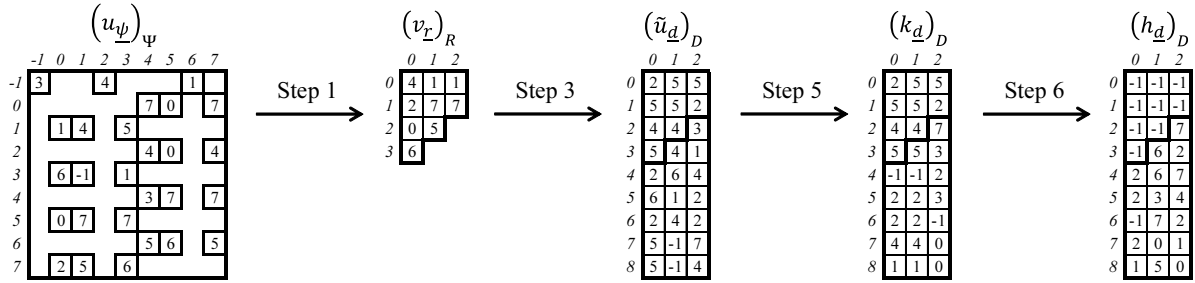}}
\caption{Numerical example of Algorithm \ref{Decoding of non-systematic codewords} for a non-systematic Hermitian codeword with erasure-and-errors. The Gr\"obner bases are shown in Example \ref{example-non-systematic-decoding-Hermitian}.
\label{figure-non-systematic-decoding-Hermitian}}
\end{figure}
(Continued from Example \ref{example-non-systematic-encoding-Hermitian})
As it can be shown that $d_{\mathrm{FR}}=7$ for $C^\perp(V_B,\Psi)$, the erasure-and-error correction can be performed by Algorithm \ref{Decoding of non-systematic codewords} if $|\Phi_1|+2|\Phi_2|<7$.
Erasure-and-error decoding of the non-systematic codeword in Fig.~\ref{figure-non-systematic-encoding-Hermitian} via Algorithm \ref{Decoding of non-systematic codewords} is described as follows.
The input of Algorithm \ref{Decoding of non-systematic codewords} consists of the received word $\left(r_{\underline{\psi}}\right)_\Psi$ in Fig.~\ref{figure-non-systematic-decoding-Hermitian} and a set $\Phi_1$ of erasure locations $\{(\alpha^6,\alpha^4),(\alpha^6,\alpha^7)\}$.
Fig.~\ref{figure-non-systematic-decoding-Hermitian} shows the values of vectors at each step in Algorithm \ref{Decoding of non-systematic codewords}.
In Step 2, the Gr\"obner basis $\mathcal{G}_{\Phi_1}$ of $Z_{\Phi_1}$ is obtained as
\begin{align*}
\mathcal{G}_{\Phi_1}=
\left\{
g^{(0)}=\alpha^2+x,\;
g^{(1)}=\alpha^2y+xy,\;
g^{(2)}=\alpha^3+\alpha^5y+y^2
\right\}.
\end{align*}
In Step 3, $\left(\widetilde{r}_{\underline{d}}\right)_D=\left(\sum_{\underline{\psi}\in\Psi}r_{\underline{\psi}}\underline{\psi}^{\underline{d}}\right)_D$ is computed, e.g., $\widetilde{r}_{(0,0)}=\sum_{\underline{\psi}\in\Psi}r_{\underline{\psi}}=\alpha^2$ according to the case $N=2$ and $a=b=0$ in Example 1.
In Step 4, the Gr\"obner basis $\mathcal{G}_{\Phi_1\cup\Phi_2}$ of $Z_{\Phi_1\cup\Phi_2}$ is obtained as
\begin{align*}
\mathcal{G}_{\Phi_1\cup\Phi_2}=
\left\{
\begin{array}{l}
g^{(0)}=\alpha x+\alpha^4x^2+x^3,\\
g^{(1)}=1+\alpha^7x+\alpha^2y+\alpha^2x^2+xy,\\
g^{(2)}=\alpha^5+\alpha^7x+\alpha^5y+\alpha^2x^2+y^2
\end{array}\right\}.
\end{align*}
If we perform Chien search for $\mathcal{G}_{\Phi_1\cup\Phi_2}$, the set $\Phi_1\cup\Phi_2$ of the erasure-and-error locations can be determined; however, the explicit set $\Phi_1\cup\Phi_2$ may not be used in our algorithm.
It can be seen in Fig.~\ref{figure-non-systematic-decoding-Hermitian} that the erasure-and-error spectrum $\left(k_{\underline{d}}\right)_D$ is generated by $\mathcal{G}_{\Phi_1\cup\Phi_2}$ from $\left(\widetilde{r}_{\underline{b}}\right)_B$ in Step 5, and $\left(k_{\underline{d}}\right)_D$ is then removed from $\left(\widetilde{r}_{\underline{d}}\right)_D$ in Step 6.
The resulting $\left(h_{\underline{d}}\right)_D$ agrees with the information given in Fig.\ \ref{figure-non-systematic-encoding-Hermitian}.
\QED
\end{example}
\begin{figure}[t!]% Fig. 6
\centering
  \resizebox{16cm}{!}{\includegraphics{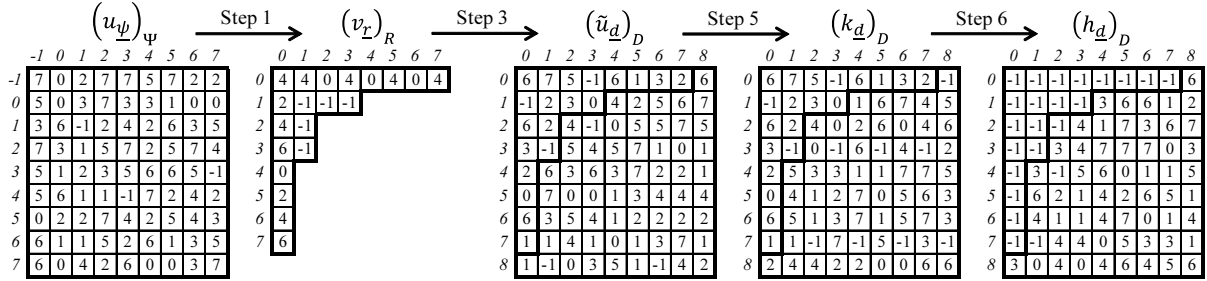}}
\caption{Numerical example of Algorithm \ref{Decoding of non-systematic codewords} for a non-systematic HCRS codeword with erasure-and-errors. The Gr\"obner bases are shown in Example \ref{example-non-systematic-decoding-HCRS}.
\label{figure-non-systematic-decoding-HCRS}}
\end{figure}
\begin{example}\label{example-non-systematic-decoding-HCRS}
(Continued from Example \ref{example-non-systematic-encoding-HCRS})
As it can be shown \cite{Hoholdt98} that $d_{\mathrm{min}}=d_{\mathrm{FR}}=9$ for $C^\perp(V_B,\Psi)$, where $d_{\mathrm{min}}$ is the true minimum distance, the erasure-and-error correction can be performed by Algorithm \ref{Decoding of non-systematic codewords} if $|\Phi_1|+2|\Phi_2|<9$.
Fig.~\ref{figure-non-systematic-decoding-HCRS} shows the data at each step of Algorithm \ref{Decoding of non-systematic codewords} for the erasure-and-error decoding of the non-systematic codeword in Fig.~\ref{figure-non-systematic-encoding-HCRS}.
The input of Algorithm \ref{Decoding of non-systematic codewords} consists of the received word $\left(r_{\underline{\psi}}\right)_\Psi$ in Fig.~\ref{figure-non-systematic-decoding-HCRS} and a set $\Phi_1$ of erasure locations $\{(0,\alpha^4),(\alpha^2,0)\}$.
Consider the graded lexicographic order \cite{Cox97} to be a monomial order $\preceq$ such that $(a_1,a_2)\preceq(a_1',a_2')\Leftrightarrow a_1+a_2<a_1'+a_2'\mbox{ or }a_1+a_2=a_1'+a_2',\,a_2\le a_2'$, i.e., $(0,0)\preceq(1,0)\preceq(0,1)\preceq(2,0)\preceq(1,1)\preceq(0,2)\preceq(3,0)\preceq\cdots\preceq(0,3)\preceq(4,0)\preceq(3,1)\preceq\cdots\preceq(8,8)$ on $A=\{0,1,\cdots,8\}^2$.
In Step 2, the Gr\"obner basis $\mathcal{G}_{\Phi_1}$ of $Z_{\Phi_1}$ is obtained as
\begin{align*}
\mathcal{G}_{\Phi_1}=
\left\{
g^{(0)}=\alpha^6x+x^2,\;
g^{(1)}=\alpha^0+\alpha^2x+y
\right\},
\end{align*}
where we denote $x=x_1$ and $y=x_2$.
In Step 4, the Gr\"obner basis $\mathcal{G}_{\Phi_1\cup\Phi_2}$ of $Z_{\Phi_1\cup\Phi_2}$ is obtained as
\begin{align*}
\mathcal{G}_{\Phi_1\cup\Phi_2}=
\left\{
\begin{array}{l}
g^{(0)}=\alpha^6+\alpha^6y+\alpha^2x^2+xy+x^3,\\
g^{(1)}=1+\alpha x+y+\alpha^5x^2+x^2y,\\
g^{(2)}=1\!+\!\alpha x\!+\!\alpha^4y\!+\!\alpha^5x^2\!+\!\alpha^3xy\!+\!y^2
\end{array}\right\}.
\end{align*}
The resulting $\left(h_{\underline{d}}\right)_D$ agrees with the information given in Fig.\ \ref{figure-non-systematic-encoding-HCRS}.
\QED
\end{example}

%%%%%%%%%%%%%%%%%%%%%%%%%%%%%%%%%%%%%%%%%%%%%%%%%%%%%%%%%%%%%%%%
\subsection{Erasure-and-error decoding: general case
\label{general case}}

In Algorithm \ref{Decoding of non-systematic codewords}, we removed the erasure-and-error spectrum from the received word spectrum without identifying $\left(e_{\underline{\psi}}\right)_\Psi$.
In this subsection, we consider the problem of erasure-and-error decoding with identifying $\left(e_{\underline{\psi}}\right)_\Psi$ in the received word.
It follows from Main Lemma that the value of $\mathcal{C}$ for the erasure-and-error spectrum is equal to $\left(e_{\underline{\psi}}\right)_\Psi$.
Though $\mathcal{F}^{-1}$ was not used in Algorithm \ref{Decoding of non-systematic codewords}, the map $\mathcal{C}$ including $\mathcal{F}^{-1}$ is required in Algorithm \ref{Finding erasures and errors}.

\begin{algorithm}{\it (Finding erasures and errors)}
\label{Finding erasures and errors}
\begin{description}
 \setlength{\itemsep}{1mm}
\item[Input:]\ $\Phi_1$ and a received word $\left(r_{\underline{\psi}}\right)_\Psi\in V_\Psi$
\item[Output:]\ \ $\left(c_{\underline{\psi}}\right)_\Psi\in C^{\perp}(V_B,\Psi)$
\item[Step 1.]\ $\left(v_{\underline{b}}\right)_B=\left(\sum_{\underline{\phi}\in\Phi_1}\underline{\phi}^{\underline{b}}\right)_B\in V_B$
\item[Step 2.]\ Calculate $\mathcal{G}_{\Phi_1}$ from syndrome $\left(v_{\underline{b}}\right)_B$
\item[Step 3.]\ $\left(\widetilde{r}_{\underline{b}}\right)_B=\left(\sum_{\underline{\psi}\in\Psi}r_{\underline{\psi}}\underline{\psi}^{\underline{b}}\right)_B\in V_B$
\item[Step 4.]\ Calculate $\mathcal{G}_{\Phi_1\cup\Phi_2}$ from $\left(\widetilde{r}_{\underline{b}}\right)_B$ and $\mathcal{G}_{\Phi_1}$
\item[Step 5.]\ $\left(e_{\underline{\psi}}\right)_\Psi=\mathcal{C}\left(\left(\widetilde{r}_{\underline{b}}\right)_B\right)\in V_\Psi$
\item[Step 6.]\ $\left(c_{\underline{\psi}}\right)_\Psi=\left(r_{\underline{\psi}}\right)_\Psi-\left(e_{\underline{\psi}}\right)_\Psi\in C^{\perp}(V_B,\Psi)$
\quad\QED
\end{description}
\end{algorithm}
\vspace{0.5em}
In this algorithm, Main Lemma is used in Step 5, because $\left(\widetilde{r}_{\underline{b}}\right)_B=\left(\sum_{\underline{\psi}\in\Psi}e_{\underline{\psi}}\underline{\psi}^{\underline{b}}\right)_B=\mathcal{C}^{-1}\left(\left(e_{\underline{\psi}}\right)_\Psi\right)$ from definition \eqref{dual code}, and $\mathcal{C}\left(\left(\widetilde{r}_{\underline{b}}\right)_B\right)=\left(e_{\underline{\psi}}\right)_\Psi$ by Main Lemma, which is applied as $\mathcal{C}:V_D\to V_{\Phi_1\cup\Phi_2}$ with $D=D(\Phi_1\cup\Phi_2)$.
Note that $\left(e_{\underline{\psi}}\right)_{\Phi_1\cup\Phi_2}=\mathcal{C}\left(\left(\widetilde{r}_{\underline{d}}\right)_D\right)$ is denoted as $\left(e_{\underline{\psi}}\right)_{\Psi}=\mathcal{C}\left(\left(\widetilde{r}_{\underline{b}}\right)_B\right)$ because $V_{\Phi_1\cup\Phi_2}\subseteq V_{\Psi}$ and $D(\Phi_1\cup\Phi_2)\subseteq B$.

\begin{figure}[t!]% Fig. 7
\centering
  \resizebox{16cm}{!}{\includegraphics{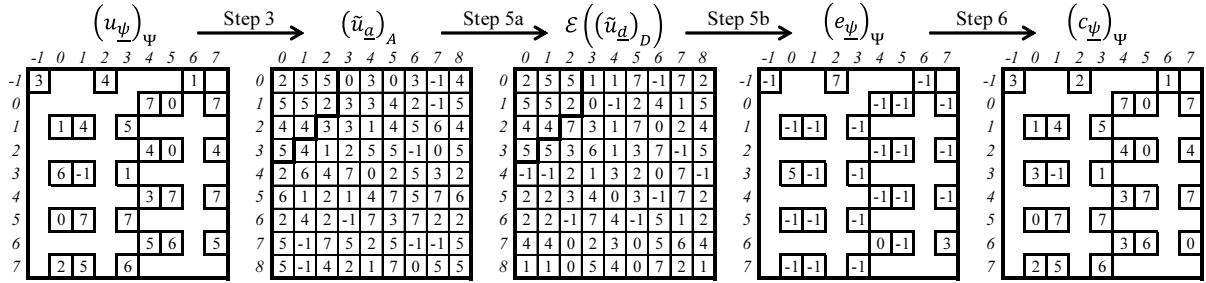}}
\caption{Numerical example of Algorithm \ref{Finding erasures and errors} for a Hermitian codeword with erasure-and-error, cf. Example \ref{example-systematic-decoding-Hermitian}. Although only $\left(\widetilde{r}_{\underline{b}}\right)_B$ is required, $\left(\widetilde{r}_{\underline{a}}\right)_A$ is shown for consistency and for the discussion in Subsection \ref{Multidimensional}. The delta set $D$ in $\mathcal{E}\left(\left(\widetilde{r}_{\underline{d}}\right)_D\right)$ indicates $D(\Phi_1\cup\Phi_2)$ and $\mathcal{E}$ is formed from $\mathcal{G}_{\Phi_1\cup\Phi_2}$.
\label{figure-systematic-decoding-Hermitian}}
\end{figure}

\begin{example}\label{example-systematic-decoding-Hermitian}
(Continued from Example \ref{example-non-systematic-decoding-Hermitian})
Erasure-and-error decoding of the codeword in Fig.~\ref{figure-non-systematic-encoding-Hermitian} via Algorithm \ref{Finding erasures and errors} is described as follows.
The input of Algorithm \ref{Finding erasures and errors} is the same as for Example \ref{example-non-systematic-decoding-Hermitian}.
Fig.~\ref{figure-systematic-decoding-Hermitian} shows the values of vectors at each step of Algorithm \ref{Finding erasures and errors}.
The Gr\"obner basis $\mathcal{G}_{\Phi_1}$ in Step 2 and the Gr\"obner basis $\mathcal{G}_{\Phi_1\cup\Phi_2}$ in Step 4 are the same as those in Example \ref{example-non-systematic-decoding-Hermitian}.
Although $\mathcal{C}$ is used in Step 5 of Algorithm \ref{Finding erasures and errors}, the value of $\mathcal{E}\left(\left(\widetilde{r}_{\underline{d}}\right)_D\right)$ is given in Fig.~\ref{figure-systematic-decoding-Hermitian} in order to show the process.
\QED
\end{example}

\begin{example}\label{example-systematic-decoding-HCRS}
\begin{figure}[t!]% Fig. 8
\centering
  \resizebox{16cm}{!}{\includegraphics{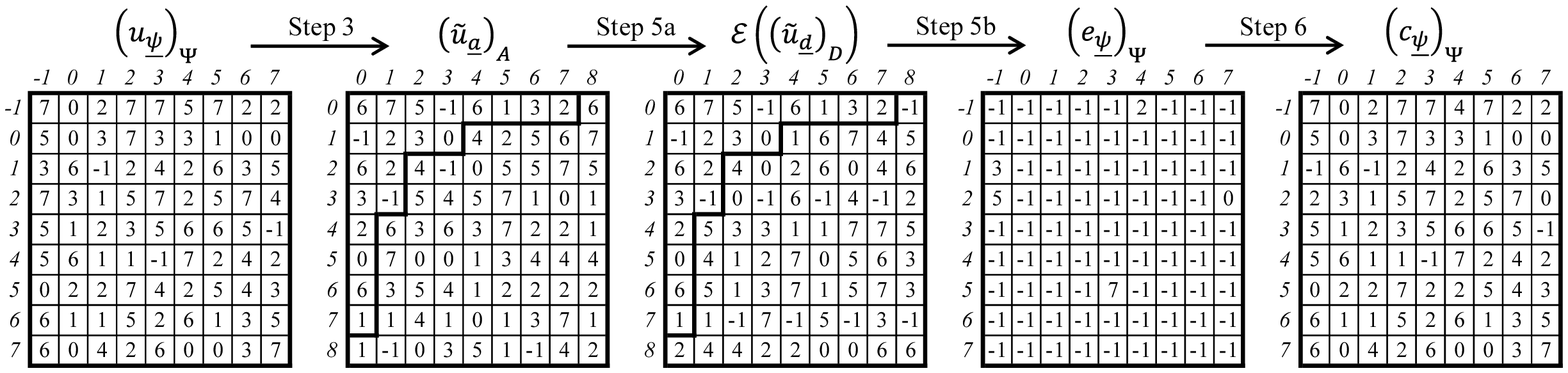}}
\caption{Numerical example of Algorithm \ref{Finding erasures and errors} for an HCRS codeword with erasure-and-error, cf. Example \ref{example-systematic-decoding-HCRS}. As noted in Example \ref{example-systematic-decoding-Hermitian}, $\mathcal{E}\left(\left(\widetilde{r}_{\underline{d}}\right)_D\right)$ is the computational process of $\mathcal{C}=\mathcal{F}^{-1}\circ\mathcal{E}$ in Step 5 of Algorithm \ref{Finding erasures and errors}. A received word $\left(r_{\underline{\psi}}\right)_\Psi$ is decomposed into $\left(e_{\underline{\psi}}\right)_\Psi$ and $\left(c_{\underline{\psi}}\right)_\Psi$.
\label{figure-systematic-decoding-HCRS}}
\end{figure}
(Continued from Example \ref{example-non-systematic-decoding-HCRS})
Erasure-and-error decoding of the codeword in Fig.~\ref{figure-non-systematic-encoding-HCRS} via Algorithm \ref{Finding erasures and errors} is described as follows.
The input of Algorithm \ref{Finding erasures and errors} is the same as for Example \ref{example-non-systematic-decoding-HCRS}.
All data at each step of Algorithm \ref{Finding erasures and errors} are shown in Fig.~\ref{figure-systematic-decoding-HCRS}.
The Gr\"obner basis $\mathcal{G}_{\Phi_1}$ in Step 2 and the Gr\"obner basis $\mathcal{G}_{\Phi_1\cup\Phi_2}$ in Step 4 are the same as those in Example \ref{example-non-systematic-decoding-HCRS}.
\QED
\end{example}

\begin{remark}
One might consider that, as the Gr\"obner basis $\mathcal{G}_{\Phi_1\cup\Phi_2}$ is obtained in Step 4 of Algorithms 1 and 2, and the set $\Phi_1\cup\Phi_2$ of erasure-and-error locations can be calculated by Chien search, the erasure-and-error values $\left(e_{\underline{\psi}}\right)_{\Phi_1\cup\Phi_2}$ can be computed from the system of linear equations $\left(\sum_{\underline{\phi}\in\Phi_1\cup\Phi_2}e_{\underline{\phi}}\underline{\phi}^{\underline{d}}\right)_D=\left(\widetilde{r}_{\underline{d}}\right)_D$ with $D=D(\Phi_1\cup\Phi_2)$, the matrix of which is invertible by \eqref{evaluation} and Appendix \ref{transpose}.
If we use Gaussian elimination to solve this, then the computational complexity is of the order $\left(\left|\Phi_1\right|+\left|\Phi_2\right|\right)^3$, which is bounded by $n^3$.
We will see in the next section that the computational complexity of Step 5 in Algorithm 1 or 2 for finding the erasure-and-error values or spectrum is bounded by the order $qn^{2+\varepsilon}$ with any $0<\varepsilon<1$.
Consequently, we can choose an appropriate method according to $\left|\Phi_1\right|+\left|\Phi_2\right|$ and $n$.
\QED
\end{remark}

%%%%%%%%%%%%%%%%%%%%%%%%%%%%%%%%%%%%%%%%%%%%%%%%%%%%%%%%%%%%%%%%
\subsection{Systematic encoding regarded as erasure-only decoding
\label{Systematic encoding}}

Because, in practical use, error-correcting codes are usually encoded systematically, it is natural to consider the systematic encoding of $C^\perp(V_B,\Psi)$.
In this subsection, we show that the systematic encoding is equivalent to a certain type of erasure-only decoding under Algorithm \ref{Finding erasures and errors}.

Systematic encoding means that there exists at least one $\Phi$ with $\Phi\subseteq\Psi$ and $|\Phi|=|B|$ such that, for any given information $\left(h_{\underline{\psi}}\right)_{\Psi\backslash\Phi}\in V_{\Psi\backslash\Phi}$, we find $\left(c_{\underline{\psi}}\right)_\Psi\in C^\perp(V_B,\Psi)$ with $c_{\underline{\psi}}=h_{\underline{\psi}}$ for all $\underline{\psi}\in\Psi\backslash\Phi$.
Thus, $\Phi$ corresponds to the set of redundant locations, and $\Psi\backslash\Phi$ corresponds to the set of information locations, in the codewords of $C^\perp(V_B,\Psi)$.
If $\Phi$ is fixed, then systematic encoding can be viewed as the erasure-only decoding of $\left(e_{\underline{\phi}}\right)_\Phi=\left(-c_{\underline{\phi}}\right)_\Phi$.
However, as $|\Phi_1|=n-k=|\Phi|$ and $|\Phi_2|=0$, the correctable erasure-and-error bound $|\Phi_1|+2|\Phi_2|<d_\mathrm{FR}$ is not generally valid.

\begin{example}\label{counterexample}
(Continued from Examples \ref{example-non-systematic-decoding-Hermitian} and \ref{example-non-systematic-decoding-HCRS})
In Examples \ref{example-non-systematic-encoding-Hermitian} and \ref{example-non-systematic-decoding-Hermitian}, because $|B|=n-k=9$ and $d_\mathrm{FR}=7$ for the Hermitian code, the correctable erasure-only bound $|B|=|\Phi_1|<d_\mathrm{FR}$ is not valid.
Similarly, in Examples \ref{example-non-systematic-encoding-HCRS} and \ref{example-non-systematic-decoding-HCRS}, because $|B|=n-k=20$ and $d_\mathrm{FR}=9$ for the extended HCRS code, the correctable erasure-only bound $|B|=|\Phi_1|<d_\mathrm{FR}$ is also not valid.
\QED
\end{example}

Nevertheless, we can show that, in many cases, there exists $\Phi$ such that the systematic encoding works as an erasure-only decoding on $\Phi$.
We now state the condition for the erasure-only decoding under Algorithm \ref{Finding erasures and errors} with $|\Phi|=|B|$.

\begin{corollary}
{\it(Erasure-only decodable condition)}
Suppose that an erasure-only $\left(e_{\underline{\psi}}\right)_\Psi$ has occurred in a received word $\left(r_{\underline{\psi}}\right)_\Psi=\left(c_{\underline{\psi}}\right)_\Psi+\left(e_{\underline{\psi}}\right)_\Psi$ from some channel, where $\left(e_{\underline{\psi}}\right)_\Psi$ is unknown, but $\Phi\subseteq\Psi$ is known and $e_{\underline{\psi}}\not=0\Rightarrow\underline{\psi}\in\Phi$.
If the linear map $\mathrm{ev}\mid_{V_B,\Phi}$ given by
\begin{equation}\label{surjective}
\mathrm{ev}\mid_{V_B,\Phi}:V_B\to V_\Phi\quad
\left[\left(h_{\underline{b}}\right)_B\mapsto
\left(\sum_{\underline{b}\in B}h_{\underline{b}}\underline{\phi}^{\underline{b}}\right)_\Phi\right]
\end{equation}
is isomorphic, then the received word $\left(r_{\underline{\psi}}\right)_\Psi$ can be decoded by Algorithm \ref{Finding erasures and errors}.
\QED
\end{corollary}
Note that this condition is equivalent to $\det\left[\underline{x}_l\left(\underline{\phi}_m\right)\right]\not=0$, where $\left.\left\{\underline{x}^{\underline{b}}\,\right|\underline{b}\in B\right\}=\left.\left\{\underline{x}_l\,\right|\,1\le l\le|B|\right\}$ and $\Phi=\left.\left\{\underline{\phi}_m\,\right|\,1\le m\le|\Phi|\right\}$ are aligned in any order, and $\underline{x}_l\left(\underline{\phi}_m\right)$ is the $(l,m)$-th entry.
This matrix is considered in Appendix \ref{transpose}.
A non-zero determinant value $\det\left[\underline{x}_l\left(\underline{\phi}_m\right)\right]\not=0$ is expected to occur with high probability $(q-1)/q$, because the values of $\det\left[\underline{x}_l\left(\underline{\phi}_m\right)\right]$ are considered to occur equally in $\mathbb{F}_q$ if we vary $\Phi\subseteq\Psi$ and $B\subseteq D(\Psi)$ randomly.
At least when $B=D(\Phi)$, this expectation is supported experimentally for Hermitian codes by \cite{O'Sullivan}, where $\Phi$ with $\det\left[\underline{x}_l\left(\underline{\phi}_m\right)\right]\not=0$ is said to be generic, and \cite{Jensen}, where such a $\Phi$ is said to be independent.
Moreover, this expectation is supported theoretically for $\mathbb{F}_q$-rational points of algebraic curves by \cite{Hansen}.

The validity of this Corollary can be described directly as follows.
Let $\Phi\subseteq\Psi\subseteq\Omega$, so that \eqref{surjective} is isomorphic.
It follows from the surjectivity of \eqref{surjective} that, for any $\underline{a}\in A\backslash B$, there exists $\left(h_{\underline{b}}\right)_B\in V_B$ such that
$
\left(\sum_{\underline{b}\in B}h_{\underline{b}}\underline{\phi}^{\underline{b}}\right)_\Phi=\left(-\underline{\phi}^{\underline{a}}\right)_\Phi.
$
We can then find $f\in\mathbb{F}_q[\underline{x}]$ such that $f\left(\underline{\phi}\right)=0$ for all $\underline{\phi}\in\Phi$; actually, $f$ is given by
$$
f=f(\underline{x})=
\underline{x}^{\underline{a}}+\sum_{\underline{b}\in B}h_{\underline{b}}\underline{x}^{\underline{b}}\in\mathbb{F}_q[\underline{x}].
$$
Because $\underline{a}\in A\backslash B$ is arbitrary, a set of polynomials $\mathcal{G}_\Phi=\left\{f^{(w)}\right\}_{0\le w<z}$ is obtained and sufficient to extend $V_B$ into $V_A$ via $\mathcal{E}$ by \eqref{E} and \eqref{generation}; $z$ can be taken as, at most, $z\le q^{N-1}$ if $|A|=q^N$.
The syndrome
$
\left(\sum_{\underline{\psi}\in\Psi}r_{\underline{\psi}}\underline{\psi}^{\underline{b}}\right)_B=\left(\sum_{\underline{\phi}\in\Phi}e_{\underline{\phi}}\underline{\phi}^{\underline{b}}\right)_B$
can then be extended into $\mathcal{E}\left(\left(\sum_{\underline{\phi}\in\Phi}e_{\underline{\phi}}\underline{\phi}^{\underline{b}}\right)_B\right)=\left(\sum_{\underline{\phi}\in\Phi}e_{\underline{\phi}}\underline{\phi}^{\underline{a}}\right)_A$ by Proposition \ref{extension}, and by function $\mathcal{R}\circ\mathcal{F}^{-1}$, we obtain $\mathcal{C}\left(\left(\sum_{\underline{\phi}\in\Phi}e_{\underline{\phi}}\underline{\phi}^{\underline{b}}\right)_B\right)=\left(e_{\underline{\phi}}\right)_\Phi$ by Main Lemma.

The computation of $\mathcal{G}_\Phi$ can be performed by the BMS algorithm; for systematic encoding, we calculate the $\mathcal{G}_\Phi$ in advance---these play the role of generator polynomials in the case of Reed--Solomon codes.
Although the following Algorithm \ref{DFT systematic encoding} is equivalent to a special case of Algorithm \ref{Finding erasures and errors} for $\Phi_1=\Phi$ and $\Phi_2=\emptyset$, we give it separately to describe systematic encoding.

\begin{algorithm}{\it (DFT systematic encoding)}
\label{DFT systematic encoding}
\begin{description}
 \setlength{\itemsep}{1mm}
\item[Input:]\ $\Phi$ and an information word $\left(h_{\underline{\psi}}\right)_{\Psi\backslash\Phi}\in V_{\Psi\backslash\Phi}$
\item[Output:]\ \ $\left(c_{\underline{\psi}}\right)_\Psi\in C^\perp(V_B,\Psi)$ with $\left(c_{\underline{\psi}}\right)_{\Psi\backslash\Phi}=\left(h_{\underline{\psi}}\right)_{\Psi\backslash\Phi}$
\item[Step 1.]\ $\left(\widetilde{r}_{\underline{b}}\right)_B=\left(\sum_{\underline{\psi}\in\Psi\backslash\Phi}h_{\underline{\psi}}\underline{\psi}^{\underline{b}}\right)_B\in V_B$
\item[Step 2.]\ $\left(k_{\underline{a}}\right)_{A}=\mathcal{E}\left(\left(\widetilde{r}_{\underline{b}}\right)_B\right)\in V_A$ by $\mathcal{G}_{\Phi}$
\item[Step 3.]\ $\left(c_{\underline{\phi}}\right)_\Phi=-\mathcal{R}\circ\mathcal{F}^{-1}\left(\left(k_{\underline{a}}\right)_A\right)\in V_\Phi$
\quad\QED
\end{description}
\end{algorithm}
\vspace{1em}

\begin{example}\label{example-systematic-encoding-Hermitian}
\begin{figure}[t!]% Fig. 9
\centering
  \resizebox{16cm}{!}{\includegraphics{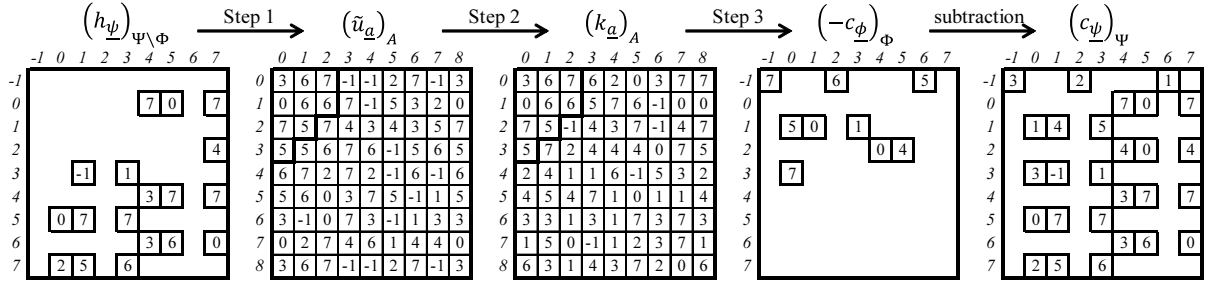}}
\caption{Numerical example of systematic encoding of the Hermitian code $C^\perp(V_B,\Psi)$ by Algorithm \ref{DFT systematic encoding}, where $\Phi$ is given by \eqref{redundant} and the Gr\"obner basis $\mathcal{G}_\Phi$ is described in Example \ref{example-systematic-encoding-Hermitian}. The given information $\left(h_{\underline{\psi}}\right)_{\Psi\backslash\Phi}$ is systematically encoded into a codeword $\left(c_{\underline{\psi}}\right)_\Psi$.
\label{figure-systematic-encoding-Hermitian}}
\end{figure}
(Continued from Example \ref{counterexample})
Let
\begin{equation}\label{redundant}
\Phi=\left\{
\begin{array}{cc}
(0,0),
(0,\alpha^2),
(0,\alpha^6),
(\alpha,1),
(\alpha,\alpha),\\
(\alpha,\alpha^3),
(\alpha^2,\alpha^4),
(\alpha^2,\alpha^5),
(\alpha^3,1)
\end{array}
\right\}.
\end{equation}
The Gr\"obner basis $\mathcal{G}_\Phi=\left\{g^{(0)},g^{(1)},g^{(2)},g^{(3)}\right\}$ is given by
\begin{align*}
&g^{(0)}=\alpha^2x+\alpha^7x^2+\alpha x^3+x^4,\\
&g^{(1)}=\alpha^7x+x^2+\alpha^4x^3+y(\alpha^3x+\alpha^4x^2+x^3),\\
&g^{(2)}=\alpha^4x^2+\alpha^7x^3+y(\alpha^4x+\alpha^7x^2)+y^2(\alpha^5x+x^2),\\
&g^{(3)}=\alpha^2x+\alpha^7x^2+\alpha x^3+y+y^3.
\end{align*}
As $D(\Phi)=B$, the isomorphy of \eqref{surjective} follows from \eqref{evaluation}.
All values of Algorithm \ref{DFT systematic encoding} are shown in Fig.\ \ref{figure-systematic-encoding-Hermitian}.
\QED
\end{example}
\begin{figure}[t!]% Fig. 10
\centering
  \resizebox{16cm}{!}{\includegraphics{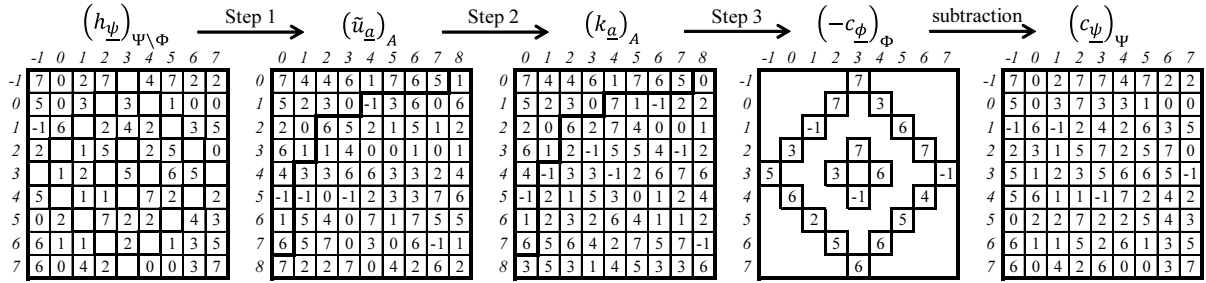}}
\caption{Numerical example of systematic encoding of the HCRS code $C^\perp(V_B,\Psi)$ by Algorithm \ref{DFT systematic encoding}, where $\Phi$ is given by \eqref{first} and the Gr\"obner basis $\mathcal{G}_\Phi$ is described in Example \ref{final}.
By regarding $\left(h_{\underline{\psi}}\right)_{\Psi\backslash\Phi}$ as a received word with erasures, the negative redundant part $\left(-c_{\underline{\phi}}\right)_{\Phi}$ is obtained.
\label{figure-systematic-HCRS}}
\end{figure}
\begin{example}\label{final}
(Continued from Example \ref{counterexample})
Let
\begin{equation}\label{first}
\Phi=\left\{
\begin{array}{c}
(0,\alpha^3),(1,\alpha^2),(1,\alpha^4),(\alpha,\alpha),(\alpha,\alpha^5),\\
(\alpha^2,1),(\alpha^2,\alpha^3),(\alpha^2,\alpha^6),(\alpha^3,0),(\alpha^3,\alpha^2),\\
(\alpha^3,\alpha^4),(\alpha^3,\alpha^7),(\alpha^4,1),(\alpha^4,\alpha^3),(\alpha^4,\alpha^6),\\
(\alpha^5,\alpha),(\alpha^5,\alpha^5),
(\alpha^6,\alpha^2),(\alpha^6,\alpha^4),
(\alpha^7,\alpha^3)
\end{array}
\right\};
\end{equation}
although we can choose $\Phi$ as
\begin{equation*}
\Phi=\left\{
\begin{array}{cc}
(0,0),
(1,0),
(\alpha,0),
(\alpha^2,0),
(\alpha^3,0),\\
(\alpha^4,0),
(\alpha^5,0),
(\alpha^6,0),
(0,1),
(1,1),\\
(\alpha,1),
(\alpha^2,1),
(0,\alpha),
(1,\alpha),
(0,\alpha^2),\\
(1,\alpha^2),
(0,\alpha^3),
(0,\alpha^4),
(0,\alpha^5),
(0,\alpha^6)
\end{array}\right\},
\end{equation*}
which has the same shape in $\Omega$ as $B\subseteq A$ because both $\Phi$ lead to $\det\left[\underline{x}_l\left(\underline{\phi}_m\right)\right]\not=0$, we adopt \eqref{first} in order to show $\Phi$'s flexibility.
The Gr\"obner basis $\mathcal{G}_\Phi=\left\{g^{(0)},g^{(1)},\cdots,g^{(8)}\right\}$ is then computed as
\begin{align*}
&g^{(0)}=\alpha^4+\alpha^2x+\alpha^2x^2+\alpha^6x^3+x^4+x^5+\alpha^6x^6+x^7\\
&+\alpha^2y+\alpha^6y^2+\alpha^6y^3+\alpha^4y^4+y^5+\alpha^2y^6+y^7+x^8,\\
&g^{(1)}=\alpha^6x+x^2+\alpha^2x^3+\alpha^5x^4+\alpha^4x^5+\alpha x^6+\alpha^3x^7\\
&+\alpha^6y+\alpha^5x^2y+\alpha x^3y+\alpha^4y^2+\alpha^5xy^2+\alpha^2y^3+\alpha^5xy^3\\
&+\alpha y^4+y^5+\alpha^5y^6+\alpha^3y^7+x^4y,\\
&g^{(2)}=
     \alpha^4 x
     +\alpha^2 x^2
     +\alpha^5 x^3
     +\alpha^7 x^5
     +\alpha^4 y
     +\alpha^6 xy
     +\alpha^6 x^2y\\
    &+\alpha^4 x^3y
     +\alpha^2 y^2
     +\alpha^6 xy^2
     +\alpha^5 y^3
     +\alpha^4 xy^3
     +\alpha^7 y^5
     + x^2y^2
,\\
&g^{(3)}=
     \alpha^6 x
     +\alpha^5 x^2
     +\alpha x^3
     +\alpha^2 x^4
     +\alpha^5 x^5
     +\alpha^2 x^6
     +\alpha^6 y\\
    &+ x^2y
     +\alpha x^3y
     +\alpha y^2
     +\alpha xy^2
     +\alpha^3 y^3
     +\alpha^5 xy^3
     +\alpha^6 y^4\\
    &+\alpha^3 y^5
     +\alpha^6 y^6
     +x^2y^3
,\\
&g^{(4)}=
     \alpha^6 x
     +\alpha^4 x^2
     +\alpha^2 x^3
     +\alpha x^4
     + x^5
     +\alpha^5 x^6
     +\alpha^3 x^7\\
    &+\alpha^6 y
     +\alpha^5 x^2y
     +\alpha^5 x^3y
     + y^2
     +\alpha^5 xy^2
     +\alpha^2 y^3
     +\alpha xy^3\\
    &+\alpha^5 y^4
     +\alpha^4 y^5
     +\alpha y^6
     +\alpha^3 y^7
     +xy^4
,\\
&g^{(5)}=
     \alpha^3 x
     +\alpha^7 x^2
     + x^3
     +\alpha^4 x^4
     +\alpha^2 x^5
     +\alpha^4 x^7
     +\alpha^3 y\\
    &+\alpha^3 x^2y
     +\alpha^3 x^3y
     +\alpha^3 y^2
     +\alpha^3 xy^2
     + y^3
     +\alpha^5 xy^3
     +\alpha y^4\\
    &+\alpha^2 y^5
     + y^6
     +\alpha^4 y^7
     +xy^5
,\\
&g^{(6)}=
     \alpha x
     +\alpha x^2
     +\alpha^2 x^3
     +\alpha^2 x^4
     +\alpha^7 x^5
     +\alpha^4 x^7
     +\alpha y\\
    &+\alpha^3 x^2y
     +\alpha^3 x^3y
     +\alpha^5 y^2
     +\alpha^2 xy^2
     +\alpha^5 y^3
     +\alpha^7 xy^3\\
    &+\alpha^6 y^4
     +\alpha^7 y^5
     + xy^6
,\\
&g^{(7)}=
     1
     +\alpha x
     +\alpha^4 x^2
     +\alpha^5 x^4
     +\alpha^6 x^6
     +\alpha^3 x^7
     +\alpha y\\
     &+\alpha^2 xy
     +\alpha^7 x^2y
     +\alpha^7 x^3y
     +\alpha^3 y^2
     +\alpha^7 xy^2
     +\alpha^3 xy^3\\
     &+\alpha y^4
     +\alpha^2 y^6
     +\alpha^3 y^7
    +xy^7
,\\
&g^{(8)}=
     \alpha^4
     +\alpha^2 x
     +\alpha^6 x^2
     +\alpha^6 x^3
     +\alpha^4 x^4
     + x^5
     +\alpha^2 x^6\\
     &+ x^7
     +\alpha^2 y
     +\alpha^2 y^2
     +\alpha^6 y^3
     + y^4
     + y^5
     +\alpha^6 y^6
     + y^7
    +y^8.
\end{align*}
All values of Algorithm \ref{DFT systematic encoding} are shown in Fig.\ \ref{figure-systematic-HCRS}.
\QED
\end{example}

Thus, the systematic encoding can be viewed as a special case of Algorithm \ref{Finding erasures and errors} for $\left(r_{\underline{\psi}}\right)_\Psi=\left(h_{\underline{\psi}}\right)_\Psi$ with $h_{\underline{\psi}}=0$ for all $\underline{\psi}\in\Phi$.
As there are many cases where the erasure-only correctable bound is exceeded, it is expected that both erasure-only and erasure-and-error can often be decoded beyond the erasure-and-error correcting bound $|\Phi_1|+2|\Phi_2|<d_\mathrm{FR}$.
In \cite{ISITA10}, the improvement and the necessary and sufficient condition for generic erasure-and-error decoding to succeed are obtained for Hermitian codes.

\begin{remark}
If linear codes have non-trivial automorphism groups, then systematic encoding can also be performed by a division algorithm via Gr\"obner bases for modules \cite{Heegard95},\cite{Little09}.
Indeed, there are cases where its computational complexity is less than that of Algorithm \ref{DFT systematic encoding}, as shown in \cite{Chen},\cite{Van}.
On the other hand, our method is more widely applicable to codes independent of automorphism groups.
Another advantage of our method is that there are cases where encoding and erasure-and-error decoding are integrated, and thereby the overall size of the encoder and decoder is reduced; for the case of Reed--Solomon codes, see \cite{Magnetics09}.
\QED
\end{remark}

%%%%%%%%%%%%%%%%%%%%%%%%%%%%%%%%%%%%%%%%%%%%%%%%%%%%%%%%%%%%%%%%
\section{Estimation of Complexity
\label{Estimation}}

%%%%%%%%%%%%%%%%%%%%%%%%%%%%%%%%%%%%%%%%%%%%%%%%%%%%%%%%%%%%%%%%
\subsection{Simple counting
\label{Simple}}

We now estimate the number of finite-field operations, i.e., additions, subtractions, multiplications, and divisions, required by our method.
We consider Algorithm \ref{Finding erasures and errors} for the code $C^\perp(V_B,\Psi)$, as our systematic encoding algorithm corresponds to a special case of Algorithm \ref{Finding erasures and errors}.
In this subsection, we simply count the operations in each step of the algorithm.

A summary of the results of our evaluation is given in Table \ref{simply}, where $n$ is the code length, $N$ is the dimension of $\Omega$, $q$ is the finite-field size, $z$ is the number of elements in the Gr\"obner bases, and Step 5 is decomposed into Step 5a of $\left(k_{\underline{a}}\right)_A=\mathcal{E}\left(\left(\widetilde{r}_{\underline{b}}\right)_B\right)$ and Step 5b of $\left(e_{\underline{\psi}}\right)_\Psi=\mathcal{R}\circ\mathcal{F}^{-1}\left(\left(k_{\underline{a}}\right)_A\right)$.
\begin{table}[t]
\caption{Number of Finite-Field Operations in Algorithm \ref{Finding erasures and errors}\label{simply}}
{\renewcommand\arraystretch{1.5}
\begin{center}
\begin{tabular}{|c|c|c|}
\hline
Algorithm \ref{Finding erasures and errors} &manipulation &order of bound\\
\hline
Step 1 &$\left(\sum_{\underline{\phi}\in\Phi_1}\underline{\phi}^{\underline{b}}\right)_B$ &$Nn^2$\\
Step 2 &BMS &$zn^2$\\
Step 3 &$\left(\sum_{\underline{\psi}\in\Psi}r_{\underline{\psi}}\underline{\psi}^{\underline{b}}\right)_B$ &$Nn^2$\\
Step 4 &BMS &$zn^2$\\
Step 5a &$\mathcal{E}\left(\left(\widetilde{r}_{\underline{b}}\right)_B\right)$ &$nq^N$\\
Step 5b &$\mathcal{R}\circ\mathcal{F}^{-1}\left(\left(k_{\underline{a}}\right)_A\right)$ &$nNq^N$\\
Step 6 &$\left(r_{\underline{\psi}}\right)_\Psi-\left(e_{\underline{\psi}}\right)_\Psi$ &$n$\\
\hline
\end{tabular}
\end{center}}
\end{table}
We now consider the above estimation of each step.

Step 1) The calculation of DFT $\sum_{\underline{\phi}\in\Phi_1}\underline{\phi}^{\underline{b}}$ can be decomposed into updating $\underline{\phi}^{\underline{b}}$ and adding to the preserved value.
This means that, at most, $N+1$ operations are repeated $|\Phi_1|$ times, so $(N+1)|\Phi_1|$ operations are required to compute one sum $\sum_{\underline{\phi}\in\Phi_1}\underline{\phi}^{\underline{b}}$.
As there are at most $|\Psi|=n$ values on $B\subseteq D(\Psi)$, the total number of $\mathbb{F}_q$-operations in Step 1 has an upper bound of the order $Nn^2$.

Step 2) The computational complexity of the BMS algorithm \cite{generic06},\cite{BMS05} is quoted as $zn^2$.

Step 3) Similarly to Step 1, the calculation of DFT $\sum_{\underline{\psi}\in\Psi}r_{\underline{\psi}}\underline{\psi}^{\underline{b}}$ can be decomposed into updating $\underline{\psi}^{\underline{b}}$, multiplying by $r_{\underline{\psi}}$, and adding to the preserved value.
As these three operations are repeated $|\Psi|$ times, $(N+2)|\Psi|$ operations are required to compute one sum $\sum_{\underline{\psi}\in\Psi}r_{\underline{\psi}}\underline{\psi}^{\underline{b}}$.
As there are at most $n$ values on $B$, the total number of $\mathbb{F}_q$-operations in Step 3 has an upper bound of the order $Nn^2$.

Step 4) The order $zn^2$ is quoted, as for Step 2.

Step 5a) For the extension of syndrome values, there are $2|D(\Psi)|=2|\Psi|$ additions and multiplications in the recurrence \eqref{generation}.
Thus, the order of the upper bound for the extension is $nq^N$.

Step 5b) Similarly to Step 3, the calculation of $\mathcal{F}^{-1}$ can be decomposed into updating $\omega_{i_1}^{-l_1}\cdots\omega_{i_m}^{-l_m}$, summing
$$\sum_{J\subseteq\{1,2,\cdots,N\}\backslash I}(-1)^{|J|}k_{\underline{i}(I,J)},$$
multiplying, and adding to the preserved value.
The total number is $\left(m+2^{N-m}+2\right)q^m$, which is bounded by $(N+3)q^N$.
As these operations are repeated $n$ times, the total number of $\mathbb{F}_q$-operations in Step 5b has an upper bound of the order $nNq^N$.

Step 6) Exactly $|\Psi|=n$ subtractions are performed.

Because $N\le z$, the total number of operations in Algorithm \ref{Finding erasures and errors} has an upper bound of the order $zn^2+nNq^N$.
If $N=1$, then we have $n\le q$ and $zn^2+nNq^N\le 2q^2$.
Suppose that $N>1$.
In the proof \cite{Fitzgerald-Lax98} of $\{\mbox{linear codes}\}=\{\mbox{affine variety codes}\}$, $q^N$ is chosen as $q^{N-1}<n\le q^N$, which leads to $q^N<qn$ and $N-1<\log_q n\le N$.
Then, $zn^2+nNq^N$ has an upper bound of the order $n^2\left(z+q\log_qn\right)$; the factor $\left(z+q\log_qn\right)$ is comparatively less than $n$.
Thus, Algorithm \ref{Finding erasures and errors} improves the order $n^3$ of the total computational complexity of the erasure-and-error decoding by the Gaussian elimination.
Our method based on Main Lemma reduces the complexity of evaluating erasure-and-error values from $O(n^3)$ to $O(n^2q\log_qn)$.

%%%%%%%%%%%%%%%%%%%%%%%%%%%%%%%%%%%%%%%%%%%%%%%%%%%%%%%%%%%%%%%%
\subsection{Application of m-D DFT algorithm
\label{Multidimensional}}

In Steps 1, 3, and 5b of Algorithm \ref{Finding erasures and errors}, the computations of DFT and IDFT are restricted to values on $B$ and $\Psi$, respectively.
In this subsection, we consider the algorithm that enlarges their computations to $A$ and $\Omega$, i.e., the algorithm that replaces Steps 1, 3, and 5b with the following.
\begin{description}
 \setlength{\itemsep}{1mm}
\item[Step 1${}'\!.$]\quad$\left(v_{\underline{a}}\right)_A=\left(\sum_{\underline{\psi}\in\Phi_1}\underline{\psi}^{\underline{a}}\right)_A\in V_A$
\item[Step 3${}'\!.$]\quad$\left(\widetilde{r}_{\underline{a}}\right)_{A}=\left(\sum_{\underline{\psi}\in\Psi}r_{\underline{\psi}}\underline{\psi}^{\underline{a}}\right)_A\in V_A$
\item[Step 5b${}'\!.$]\quad$\left(e_{\underline{\omega}}\right)_\Omega=\mathcal{F}^{-1}\left(\left(k_{\underline{a}}\right)_A\right)\in V_\Omega$
\end{description}
If the complexity of Steps 1${}'$, 3${}'$, and 5b${}'$ is estimated by the same method as for Steps 1, 3, and 5b, the result is an upper bound of the order $Nq^{2N}$.
It is well-known that the computational complexity of the ordinary FFT is of the order $L\log L$, where $L$ is the size of the data.
As $L=q^N$ in our case, $L\log L$ is equal to $Nq^N\log q$, though the ordinary FFT cannot be applied to our DFT and IDFT over the finite field.
By applying the inductive expressions in Section \ref{Properties}, we find the computational complexities of Steps 1${}'$, 3${}'$, and 5b${}'$ to be as shown in Table \ref{modify}.
\begin{table}[t]
\caption{Number of Finite-Field Operation in Algorithm \ref{Finding erasures and errors} Applied an m-D DFT Algorithm to Steps 1, 3, and 5{\rm b}\label{modify}}
\begin{center}
{\renewcommand\arraystretch{1.3}
\begin{tabular}{|c|c|c|}
\hline
Algorithm \ref{Finding erasures and errors} & manipulation & order of bound \\
\hline
Step 1${}'$ & $\left(\sum_{\underline{\phi}\in\Phi_1}\underline{\phi}^{\underline{a}}\right)_A$ & $Nq^{N+1}$\\
Step 3${}'$ & $\left(\sum_{\underline{\psi}\in\Psi}r_{\underline{\psi}}\underline{\psi}^{\underline{a}}\right)_{A}$ & $Nq^{N+1}$\\
Step 5b${}'$ & $\mathcal{F}^{-1}\left(\left(k_{\underline{a}}\right)_A\right)$ &
$Nq^{N+1}$\\
\hline
\end{tabular}}
\end{center}
\end{table}

Because of Propositions \ref{low-dim} and \ref{lower-dim}, we can argue DFT and IDFT identically, and focus on DFT.
It is shown by induction that the computational complexity of calculating $\mathcal{F}\left(\left(c_{\underline{\omega}}\right)_\Omega\right)=\left(\sum_{\underline{\omega}\in\Omega}c_{\underline{\omega}}\underline{\omega}^{\underline{a}}\right)_A$ is bounded by $3Nq^{N+1}$.
For $N=1$, we obtain the bound $3q^{2}$, as $\sum_{\omega\in\Omega}c_{\omega}\omega^a$ is decomposed into updating $\omega^a$, multiplying by $c_\omega$, and adding to the preserved value for all $\omega\in\Omega=\mathbb{F}_q$ and for all $a\in A=\{0,1,\cdots,q-1\}$.
Assume that, for $N-1$, we obtain the bound $3(N-1)q^N$.
The summation can be decomposed as
\begin{equation}\label{decomposition}
\sum_{\omega_N\in\mathbb{F}_q}\omega_N^{a_N}
\sum_{\left(\omega_1,\cdots,\omega_{N-1}\right)\in\mathbb{F}_q^{N-1}}c_{\left(\omega_1,\cdots,\omega_{N-1},\omega_{N}\right)}\omega_1^{a_1}\cdots\omega_{N-1}^{a_{N-1}}.
\end{equation}
By induction hypothesis, the complexity of the interior summation in \eqref{decomposition} for all $a_1,\cdots,a_{N-1}\in\{0,1,\cdots,q-1\}$ is bounded by $3(N-1)q^N$.
For all $\omega_N\in\mathbb{F}_q$, the values of the interior summation are calculated in advance.
The complexity of the exterior summation in \eqref{decomposition} for all $a_N\in\{0,1,\cdots,q-1\}$ is then bounded by $3q^2$, from the case of $N=1$.
As the exterior summation is carried out for all $a_1,\cdots,a_{N-1}\in\{0,1,\cdots,q-1\}$, the total complexity of computing $\mathcal{F}\left(\left(c_{\underline{\omega}}\right)_\Omega\right)$ is bounded by
$$
3(N-1)q^N\times q+3q^2\times q^{N-1}=3Nq^{N+1}.
$$
Thus, all DFT and IDFT parts of Algorithm \ref{Finding erasures and errors} are bounded by the order $Nq^{N+1}$.
On the basis of the inductive expressions, the order $nNq^N$ in the previous subsection is changed to the order $Nq^{N+1}$, where the factor $n$ in $nNq^N$ is reduced to $q$.

Finally, we show that the complexity $O(n^2q\log_qn)$ of evaluating erasure-and-error values using the Main Lemma is improved by m-D DFT algorithm to $O(qn^2)$.
It follows from $q^N<qn$ that the complexity of order $nq^N$ for Step 5a is bounded by $qn^2$.
Moreover, from $q^{N-1}<n\le q^N$, the complexity of order $Nq^{N+1}$ for DFT and IDFT is
$$
Nq^{N+1}<\left(1+\log_qn\right)q\cdot qn,
$$
where the factor $\left(1+\log_qn\right)q$ is generally much lower than $n$.
Strictly, we have $\left(\log_qn\right)q\le n$ for $n\ge q$ and $q\ge3$; if $q=2$, then $\left(\log_qn\right)q\le n$ is valid except for $2\le n\le 4$.
Thus, the m-D DFT algorithm improves the complexity $O(n^2q\log_qn)$ of evaluating erasure-and-error values to $O(qn^2)$.

In the above estimation, the order $nq^N$ for Step 5a is dominant in $O(qn^2)$.
However, note that the equality \eqref{generation} that defines the extension $\mathcal{E}$ is almost identical to that of the discrepancy of the BMS algorithm.
Actually, in the BMS algorithm, the discrepancy $\mathcal{D}_{\underline{a}}\left(g^{(w)}\right)$ of updating polynomial $g^{(w)}\in\mathbb{F}_q[\underline{x}]$ at $\underline{a}\in A$ is represented by
$$
\mathcal{D}_{\underline{a}}\left(g^{(w)}\right)=
k_{\underline{a}}+\sum_{\underline{d}\in D(\Psi)}
g_{\underline{d}}^{(w)}k_{\underline{a}+\underline{d}-\underline{a}_w},
$$
for which the summation is the same as in \eqref{generation}.
Thus, the computation of $\mathcal{E}\left(\left(\widetilde{r}_{\underline{b}}\right)_B\right)$ in Step 5a can be considered as the extended part of the BMS algorithm, and does not cause serious damage in practice.

%%%%%%%%%%%%%%%%%%%%%%%%%%%%%%%%%%%%%%%%%%%%%%%%%%%%%%%%%%%%%%%%
\section{Conclusion
\label{Conclusion}}

Conventionally, the m-D DFT and IDFT over $\mathbb{F}_q$ are seen as transforms between two vector spaces, each of which is indexed by $\left(\mathbb{F}_q^\times\right)^N$.
In this paper, we have generalized these to transforms between two vector spaces, each of which is indexed by $\mathbb{F}_q^N$.
Moreover, the Fourier inversion formulae of their transforms has also been generalized.
We obtained a lemma using the linear recurrence relations from Gr\"obner bases and the generalized inverse transforms.
This states that there is a canonical one-to-one linear map from a vector space indexed by the delta set of Gr\"obner bases onto another vector space indexed by an arbitrary subset of $\mathbb{F}_q^N$.
As an application of our lemma, we have described the construction of affine variety codes, and have shown that the systematic encoding of a class of dual affine variety codes is nothing but a special case of erasure-only decoding.
As another application of our lemma, we have proposed a fast error-value estimation in the erasure-and-error decoding of the class of dual affine variety codes.
We have improved the computational complexity of the error-value estimation from $O(n^3)$ under Gaussian elimination to $O(qn^2)$, where $n$ is the code length.
Because error-value estimation with Gaussian elimination affects the speed of the BMS algorithm, we have accomplished the fast decoding of dual affine variety codes only after the Main Lemma has been used for error-value estimation.
Future work will concentrate on improving the error-correcting capability of generic erasure-and-error cases.

\appendices
%%%%%%%%%%%%%%%%%%%%%%%%%%%%%%%%%%%%%%%%%%%%%%%%%%%%%%%%%%%%%%%%
\section{Proof of Proposition \ref{Fourier}
\label{Proof of Fourier}}

It may be proved that, for $\left(c'_{\underline{\omega}}\right)_\Omega\in V_\Omega$, if $\left(h_{\underline{a}}\right)_A=\mathcal{F}\left(\left(c'_{\underline{\omega}}\right)_\Omega\right)$ and $\left(c_{\underline{\omega}}\right)_\Omega=\mathcal{F}^{-1}\left(\left(h_{\underline{a}}\right)_A\right)$ are defined, then $\left(c_{\underline{\omega}}\right)_\Omega=\left(c'_{\underline{\omega}}\right)_\Omega$ holds.
\footnote{If $\mathcal{F}^{-1}\circ\mathcal{F}=\mathrm{id.}$, then $\mathcal{F}$ is injective and $\mathcal{F}^{-1}$ is surjective, and it follows from $\dim V_A=\dim V_\Omega$ that $\mathcal{F}$ and $\mathcal{F}^{-1}$ are isomorphic and that $\mathcal{F}\circ\mathcal{F}^{-1}=\mathrm{id.}$.}
Hence, we will show that, for all $\underline{\omega}\in\Omega$, $c_{\underline{\omega}}=c'_{\underline{\omega}}$.
Let $I=I_{\underline{\omega}}=\{i_1,\cdots,i_m\}$ be as in Definition \ref{inverse-Fourier}.
We denote $\overline{\omega}=(-1)^m\omega_{i_1}^{-l_1}\cdots\omega_{i_m}^{-l_m}$.
First, note that
\begin{align}\nonumber
&\quad c_{\underline{\omega}}
=\sum_{l_1,\cdots,l_m=1}^{q-1}
\left\{
\sum_{J\subseteq\{1,2,\cdots,N\}\backslash I}(-1)^{|J|}h_{\underline{i}(I,J)}
\right\}\overline{\omega}\qquad\mbox{($\because$ Definition \ref{inverse-Fourier})}\\
&=\sum_{l_1,\cdots,l_m=1}^{q-1}\nonumber
\left\{
\sum_{J\subseteq\{1,2,\cdots,N\}\backslash I}(-1)^{|J|}\sum_{\underline{\psi}\in\Omega}c'_{\underline{\psi}}\underline{\psi}^{\underline{i}(I,J)}
\right\}\overline{\omega}\qquad\mbox{($\because$ assumption and Definition \ref{Generalization})}\\
&=\sum_{l_1,\cdots,l_m=1}^{q-1}\label{interior}
\left\{\sum_{\underline{\psi}\in\Omega}c'_{\underline{\psi}}\left(
\sum_{J\subseteq\{1,2,\cdots,N\}\backslash I}(-1)^{|J|}\underline{\psi}^{\underline{i}(I,J)}\right)
\right\}\overline{\omega}.\qquad\mbox{($\because$ changing the order of sums)}
\end{align}
Next, we compute the most interior sum in \eqref{interior}.
It follows immediately from the development that
\footnote{
The equality \eqref{inclusion-exclusion} is also known as a variant of the inclusion-exclusion principal \cite{Matousek-Nesetril}.}
\begin{equation}\label{inclusion-exclusion}
\prod_{i\in\{1,2,\cdots,N\}\backslash I_{\underline{\omega}}}\left(1-\psi_i^{q-1}\right)
=\sum_{J\subseteq\{1,2,\cdots,N\}\backslash I_{\underline{\omega}}}(-1)^{|J|}\prod_{j\in J}\psi_j^{q-1}.
\end{equation}
On the other hand, because $\psi_i^{q-1}=1\Longleftrightarrow\psi\not=0$, we have
\begin{align}\label{characteristic}
\prod_{i\in\{1,2,\cdots,N\}\backslash I_{\underline{\omega}}}\left(1-\psi_i^{q-1}\right)
&=\left\{
\begin{array}{cl}
1 & \mbox{if }\psi_i=0\mbox{ for }\forall\,i\in\{1,2,\cdots,N\}\backslash I_{\underline{\omega}}\\
0 & \mbox{if }\exists\,i\in\{1,2,\cdots,N\}\backslash I_{\underline{\omega}}\mbox{ with }\psi_i\not=0.
\end{array}\right.
\end{align}
Then, the value $\displaystyle\sum\limits_{J\subseteq\{1,2,\cdots,N\}\backslash I_{\underline{\omega}}}(-1)^{|J|}\prod_{j\in J}\psi_j^{q-1}$ is equal to 1 or 0 according to the condition of \eqref{characteristic}.
Moreover, it follows from Definition \ref{inverse-Fourier} that $\underline{\psi}^{\underline{i}(I_{\underline{\omega}},J)}=\psi_{i_1}^{l_1}\cdots\psi_{i_m}^{l_m}\prod_{j\in J}\psi_j^{q-1}$.
Thus, we have, for a given $\underline{\omega}\in\Omega$ and for all $\underline{\psi}\in\Omega$,
\begin{align*}
\sum_{J\subseteq\{1,2,\cdots,N\}\backslash I_{\underline{\omega}}}(-1)^{|J|}\underline{\psi}^{\underline{i}(I_{\underline{\omega}},J)}
&=\psi_{i_1}^{l_1}\cdots\psi_{i_m}^{l_m}\sum_{J\subseteq\{1,2,\cdots,N\}\backslash I_{\underline{\omega}}}(-1)^{|J|}\prod_{j\in J}\psi_j^{q-1}\\
&=\left\{
\begin{array}{cl}
\psi_{i_1}^{l_1}\cdots\psi_{i_m}^{l_m} & \mbox{if }\psi_i=0\mbox{ for }\forall\,i\in\{1,2,\cdots,N\}\backslash I_{\underline{\omega}}\\
0 & \mbox{if }\exists\,i\in\{1,2,\cdots,N\}\backslash I_{\underline{\omega}}\mbox{ with }\psi_i\not=0.
\end{array}\right.
\end{align*}
Hence, we obtain
\begin{align*}
c_{\underline{\omega}}&=\sum_{l_1,\cdots,l_m=1}^{q-1}
\left\{
\sum_{\underline{\psi}\in\Omega,\,i\not\in I\Rightarrow\psi_i=0}c'_{\underline{\psi}}
\psi_{i_1}^{l_1}\cdots\psi_{i_m}^{l_m}
\right\}\overline{\omega},
\end{align*}
where the inner sum runs over all $\underline{\psi}\in\Omega$ which satisfies $\psi_i=0$ for all $i\in\{1,2,\cdots,N\}\backslash I_{\underline{\omega}}$.
This condition ``$i\not\in I_{\underline{\omega}}\Longrightarrow\psi_i=0$'' of $\underline{\psi}\in\Omega$ is equivalent to ``$\psi_i\not=0\Longrightarrow i\in I_{\underline{\omega}}$.''
Conversely, $\underline{\psi}\in\Omega$ with $\psi_i=0$ for some $i\in I_{\underline{\omega}}$ is not contributed to the inner sum because of the factor $\psi_{i_1}^{l_1}\cdots\psi_{i_m}^{l_m}$.
Hence, we obtain
\begin{align*}
c_{\underline{\omega}}&=\sum_{l_1,\cdots,l_m=1}^{q-1}
\left\{
\sum_{\underline{\psi}\in\Omega,\,i\in I\Leftrightarrow\psi_i\not=0}c'_{\underline{\psi}}
\psi_{i_1}^{l_1}\cdots\psi_{i_m}^{l_m}
\right\}\overline{\omega},
\end{align*}
where the inner sum runs over all $\underline{\psi}\in\Omega$ which satisfies $\psi_i=0$ for all $i\in\{1,2,\cdots,N\}\backslash I_{\underline{\omega}}$ and $\psi_i\not=0$ for all $i\in I_{\underline{\omega}}$.
\\
\indent
Finally, we change the order of the summations into
$$
c_{\underline{\omega}}=
(-1)^m\sum_{\underline{\psi}\in\Omega,\,i\in I\Leftrightarrow\psi_i\not=0}c'_{\underline{\psi}}
\sum_{l_1=1}^{q-1}\left(\frac{\psi_{i_1}}{\omega_{i_1}}\right)^{l_1}
\cdots
\sum_{l_m=1}^{q-1}\left(\frac{\psi_{i_m}}{\omega_{i_m}}\right)^{l_m},
$$
and, because $\sum_{l=1}^{q-1}\left(\psi_i/\omega_i\right)^l=0$ if $\psi_i\not=\omega_i$ and $q-1$ otherwise, we obtain $c_{\underline{\omega}}=(-1)^m(q-1)^mc'_{\underline{\omega}}=c'_{\underline{\omega}}$.
\QED

%%%%%%%%%%%%%%%%%%%%%%%%%%%%%%%%%%%%%%%%%%%%%%%%%%%%%%%%%%%%%%%%
\section{Proof of Proposition \ref{transposed}
\label{transpose}}

Consider an $n\times n$ matrix $\left[\underline{x}_l\left(\underline{\psi}_m\right)\right]$ whose $(l,m)$-th entry is equal to $\underline{x}_l\left(\underline{\psi}_m\right)$, where $\left.\left\{\underline{x}^{\underline{d}}\,\right|\,\underline{d}\in D(\Psi)\right\}=\left.\left\{\underline{x}_l\,\right|\,1\le l\le n\right\}$ and $\Psi=\left.\left\{\underline{\psi}_m\,\right|\,1\le m\le n\right\}$ are aligned by any order with $n=|\Psi|=\left|D(\Psi)\right|$.
The map $\mathrm{ev}:V_D\to V_\Psi$ of \eqref{ev} can then be represented as
\begin{equation}\label{matrix-representation 1}
\left(h_{\underline{d}}\right)_D\mapsto\left(\sum_{\underline{d}\in D}h_{\underline{d}}\underline{\psi}^{\underline{d}}\right)_\Psi
\Longleftrightarrow
\left(h_l\right)\mapsto\left(h_l\right)\left[\underline{x}_l\left(\underline{\psi}_m\right)\right],
\end{equation}
where $\left(h_l\right)$ represents any row vector of length $n$.
Moreover, the map $\mathcal{P}:V_\Psi\to V_D$ of \eqref{partial DFT} can be represented as
\begin{equation}\label{matrix-representation 2}
\left(c_{\underline{\psi}}\right)_\Psi\mapsto\left(\sum_{\underline{\psi}\in\Psi}c_{\underline{\psi}}\underline{\psi}^{\underline{d}}\right)_D
\Longleftrightarrow
\left(c_l\right)\mapsto\left(c_l\right)\left[\underline{x}_m\left(\underline{\psi}_l\right)\right],
\end{equation}
where $\left(c_l\right)$ represents any row vector of length $n$.
These facts lead to Proposition \ref{transposed} in case of the standard bases because $\left[\underline{x}_m\left(\underline{\psi}_l\right)\right]$ indicates the transpose matrix whose $(l,m)$-th entry is equal to $\underline{x}_m\left(\underline{\psi}_l\right)$.

Suppose that $\{v_1,\cdots,v_n\}$ and $\{v'_1,\cdots,v'_n\}$ are any two normal orthogonal bases of $V_D$ that consist of row vectors.
Then there exists an $n\times n$ matrix $A$ with ${}^t\!(A^{-1})=A$ such that $[v_i]=A[v'_i]$, where $[v_i]$ and $[v'_i]$ represent the matrices whose $i$-th row are equal to $v_i$ and $v'_i$ for all $1\le i\le n$.
Similarly, suppose that $\{w_1,\cdots,w_n\}$ and $\{w'_1,\cdots,w'_n\}$ are any two normal orthogonal bases of $V_\Psi$ that consist of row vectors.
Then there exists an $n\times n$ matrix $B$ with ${}^t\!B=B^{-1}$ such that $[w_i]=B[w'_i]$.
Thus the conditions \eqref{matrix-representation 1} and \eqref{matrix-representation 2} indicate that $(h_l)\left[\mathrm{ev}(v_i)\right]=(h_l)X$ and $(c_l)\left[\mathcal{P}(w_i)\right]=(c_l){}^t\!X$ with $X=\left[\underline{x}_l\left(\underline{\psi}_m\right)\right]$ if we take the standard bases $\{v_i\}$ and $\{w_i\}$ with the $n$-th identity matrix $[v_i]=[w_i]$.
Suppose that $(h_l)\left[\mathrm{ev}(v'_i)\right]=(h_l)Y[w'_i]$ and $(c_l)\left[\mathcal{P}(w'_i)\right]=(c_l)Z[v'_i]$.
Because $\left[\mathrm{ev}(v'_i)\right]=A^{-1}\left[\mathrm{ev}\left(v_i\right)\right]$ and $\left[\mathcal{P}(w'_i)\right]=B^{-1}\left[\mathcal{P}\left(w_i\right)\right]$ follow from $[v'_i]=A^{-1}[v_i]$, $[w'_i]=B^{-1}[w_i]$, and the linearity of $\mathrm{ev}$ and $\mathcal{P}$, we have
\begin{align*}
&YB^{-1}=\left[\mathrm{ev}(v'_i)\right]=A^{-1}\left[\mathrm{ev}\left(v_i\right)\right]=A^{-1}X\\
\mbox{and }\;&ZA^{-1}=\left[\mathcal{P}(w'_i)\right]=B^{-1}\left[\mathcal{P}\left(w_i\right)\right]=B^{-1}{}^t\!X.
\end{align*}
Thus we have $Y=A^{-1}XB$, $Z=B^{-1}{}^t\!XA$, and ${}^t\!Y=Z$, which leads to Proposition \ref{transposed} in case of any normal orthogonal bases.
\QED

%%%%%%%%%%%%%%%%%%%%%%%%%%%%%%%%%%%%%%%%%%%%%%%%%%%%%%%%%%%%%%%%
\section{Proof of Proposition \ref{extension}
\label{Proof of Extension}}
From the assumption, we have
\begin{equation}\label{power}
h_{\underline{d}}=\sum_{\underline{\psi}\in\Psi}c_{\underline{\psi}}\underline{\psi}^{\underline{d}},
\end{equation}
where $\underline{d}\in D=D(\Psi)$.
We now show that \eqref{power} holds not only for $\underline{d}\in D$ but also for $\underline{d}\in A$ if $\left(h_{\underline{a}}\right)_A=\mathcal{E}\left(\left(h_{\underline{d}}\right)_D\right)$.
Actually, we have
\begin{align*}
h_{\underline{a}}&=\sum_{\underline{d}\in D}v_{\underline{d}}h_{\underline{d}}
=\sum_{\underline{d}\in D}v_{\underline{d}}\sum_{\underline{\psi}\in\Psi}c_{\underline{\psi}}\underline{\psi}^{\underline{d}}\\
&=\sum_{\underline{\psi}\in\Psi}c_{\underline{\psi}}\sum_{\underline{d}\in D}v_{\underline{d}}\underline{\psi}^{\underline{d}}
=\sum_{\underline{\psi}\in\Psi}c_{\underline{\psi}}\underline{\psi}^{\underline{a}},
\end{align*}
where we use $\sum_{\underline{d}\in D}v_{\underline{d}}\underline{\psi}^{\underline{d}}=\underline{\psi}^{\underline{a}}$, which follows from \eqref{reduction} and $g^{(w)}(\underline{x})\in Z_\Psi$.
Thus, \eqref{power} holds for all $\underline{a}\in A$; in other words, $\left(h_{\underline{a}}=\sum_{\underline{\psi}\in\Psi}c_{\underline{\psi}}\underline{\psi}^{\underline{a}}\right)_A$ holds.
\QED

%%%%%%%%%%%%%%%%%%%%%%%%%%%%%%%%%%%%%%%%%%%%%%%%%%%%%%%%%%%%%%%%
\section{Proof of Proposition \ref{recurrence}
\label{Proof of recurrence}}
Let $\left(c_{\underline{\psi}}\right)_\Psi\in V_\Psi$ be $\left(h_{\underline{d}}\right)_D=\left(\sum_{\underline{\psi}\in\Psi}c_{\underline{\psi}}\underline{\psi}^{\underline{d}}\right)_D\in V_D$ according to the isomorphism $\mathcal{P}:V_\Psi\to V_D$ of \eqref{partial DFT}.
If $\left(h_{\underline{a}}\right)_A=\mathcal{E}\left(\left(h_{\underline{d}}\right)_D\right)$, then we obtain $\left(h_{\underline{a}}=\sum_{\underline{\psi}\in\Psi}c_{\underline{\psi}}\underline{\psi}^{\underline{a}}\right)_A$ by Proposition \ref{extension}.
On the other hand, because $\underline{x}^{\underline{a}-\underline{a}_w}g^{(w)}\in Z_\Psi$, we have, for all $\underline{\psi}\in\Psi$,
$$
\underline{\psi}^{\underline{a}}=-\sum_{\underline{d}\in D}
g_{\underline{d}}^{(w)}\underline{\psi}^{\underline{a}+\underline{d}-\underline{a}_w}.
$$
Thus, we have
\begin{align*}
h_{\underline{a}}&=\sum_{\underline{\psi}\in\Psi}c_{\underline{\psi}}\left\{-\sum_{\underline{d}\in D}g_{\underline{d}}^{(w)}\underline{\psi}^{\underline{a}+\underline{d}-\underline{a}_w}\right\}\\
&=-\sum_{\underline{d}\in D}g_{\underline{d}}^{(w)}\left\{\sum_{\underline{\psi}\in\Psi}c_{\underline{\psi}}\underline{\psi}^{\underline{a}+\underline{d}-\underline{a}_w}\right\}\\
&=-\sum_{\underline{d}\in D}g_{\underline{d}}^{(w)}h_{\underline{a}+\underline{d}-\underline{a}_w}.
\end{align*}

Conversely, suppose that $\left(h_{\underline{a}}\right)_A\in V_A$ satisfies that, for each $\underline{a}\in A\backslash D$, there exists at least one $0\le w<z$ such that \eqref{generation}.
If there are $\underline{a}\in A$ and $0\le v\not=w<z$ such that $\underline{a}\ge\underline{a}_v$ and $\underline{a}\ge\underline{a}_w$, then it follows from $\underline{x}^{\underline{a}-\underline{a}_v}g^{(v)}-\underline{x}^{\underline{a}-\underline{a}_w}g^{(w)}\in Z_\Psi$ that $\sum_{\underline{d}\in D}g_{\underline{d}}^{(v)}h_{\underline{a}+\underline{d}-\underline{a}_v}=\sum_{\underline{d}\in D}g_{\underline{d}}^{(w)}h_{\underline{a}+\underline{d}-\underline{a}_w}$ by the same argument as above.
Thus, $h_{\underline{a}}$ in the left-hand side of \eqref{generation} does not depend on the choice and order of $v,w$.
\QED

%%%%%%%%%%%%%%%%%%%%%%%%%%%%%%%%%%%%%%%%%%%%%%%%%%%%%%%%%%%%%%%%
\section{Proof of $C^\perp(U,\Psi)=\mathcal{C}\left(U^\perp\right)$
\label{Proof of orthogonal complement}}

We show that, for all $\left(c'_{\underline{\psi}}\right)_\Psi\in\mathrm{ev}(U)$ and all $\left(c_{\underline{\psi}}\right)_\Psi\in\mathcal{C}\left(U^\perp\right)$, the value of the inner product $\sum_{\underline{\psi}\in\Psi}c'_{\underline{\psi}}c_{\underline{\psi}}$ is equal to zero.
Let $\left(c'_l\right)=\left(c'_l\right)_{1\le l\le n}$ be the aligned $\left(c'_{\underline{\psi}}\right)_\Psi$, as in Appendix \ref{transpose}.
By \eqref{matrix-representation 1}, $\left(c'_{\underline{\psi}}\right)_\Psi\in\mathrm{ev}(U)$ is represented as
$$
\left(c'_l\right)=\left(h'_l\right)\left[\underline{x}_l\left(\underline{\psi}_m\right)\right]
\mbox{ for some }\left(h'_l\right)\in U.
$$
Similarly, let $\left(c_l\right)=\left(c_l\right)_{1\le l\le n}$ be the aligned $\left(c_{\underline{\psi}}\right)_\Psi$.
As $\mathcal{C}=\mathcal{P}^{-1}$ and \eqref{matrix-representation 2}, $\left(c_{\underline{\psi}}\right)_\Psi\in\mathcal{C}\left(U^\perp\right)$ is represented as
$$
\left(c_l\right)=\left(h_l\right)\left[\underline{x}_m\left(\underline{\psi}_l\right)\right]^{-1}
\mbox{ for some }\left(h_l\right)\in U^\perp.
$$
Because the transpose of the row vector $\left(c'_l\right)$ is equal to a column vector $\left(c'_m\right)=\left[\underline{x}_m\left(\underline{\psi}_l\right)\right]\left(h'_m\right)$, $\sum_{\underline{\psi}\in\Psi}c'_{\underline{\psi}}c_{\underline{\psi}}$ is equal to
$$
(c_l)(c'_m)=\left(h_l\right)\left[\underline{x}_m\left(\underline{\psi}_l\right)\right]^{-1}\left[\underline{x}_m\left(\underline{\psi}_l\right)\right]\left(h'_m\right)=0.\;\,\QED
$$

%%%%%%%%%%%%%%%%%%%%%%%%%%%%%%%%%%%%%%%%%%%%%%%%%%%%%%%%%%%%%%%%
\begin{table}[t]
{\renewcommand\arraystretch{0.9}
\begin{center}
\caption{List of main notation\label{Notation list}}
\begin{tabular}{|l|l|l|}
\hline
$d_{\mathrm{min}}$ & minimum distance of the code & \ref{Introduction},\ref{non-systematic case}\\
$d_{\mathrm{FR}}$ & Feng--Rao bound of $d_{\mathrm{min}}$ & \ref{Introduction},\ref{non-systematic case}\\
$n$ & code length & \ref{Introduction}\\
$q$ & finite-field size & \ref{Introduction}\\
$z$ & size of Gr\"obner basis & \ref{Introduction}\\
$N$ & dimension of index set & \ref{Introduction}\\
$D=D(\Psi)$ & delta set & \ref{Introduction},\,\ref{vector spaces}\\
$\Psi$ & subset of $\mathbb{F}_q^N$ with size $n>0$ & \ref{Introduction},\,\ref{vector spaces}\\
$\mathbb{N}_0$ & set of non-negative integers & \ref{Notation}\\
$A\backslash B$ & $\{u\in A\,|\,u\not\in B\}$ for sets $A,B$ & \ref{Notation}\\
$|S|$ & number of elements in $S$ & \ref{Notation}\\
$V_S$ & $\left.\left\{\left(v_s\right)_S\,\right|\,s\in S,\,v_s\in\mathbb{F}_q\right\}$ & \ref{Notation}\\
$A=A_N$ & $\{0,1,\cdots,q-1\}^N$ in \eqref{A} & \ref{Definitions}\\
$\Omega=\Omega_N$ & $\mathbb{F}_q^N$ in \eqref{omg} & \ref{Definitions}\\
$\underline{a}$ & an element of $A$ in \eqref{A} & \ref{Definitions}\\
$\left(h_{\underline{a}}\right)_A$ & a vector in $V_A$ & \ref{Definitions}\\
$\underline{\omega}$ & an element of $\Omega$ in \eqref{omg}& \ref{Definitions}\\
$\left(c_{\underline{\omega}}\right)_{\Omega}$ & a vector in $V_\Omega$ & \ref{Definitions}\\
$\mathcal{F}=\mathcal{F}_N$ & $N$-D generalized DFT \eqref{DFT} & \ref{Definitions}\\
$\underline{\omega}^{\underline{a}}$ & $\omega_1^{a_1}\cdots\omega_N^{a_N}$ & \ref{Definitions}\\
$\mathcal{F}^{-1}=\mathcal{F}_N^{-1}$ & $N$-D generalized IDFT \eqref{inverse},\eqref{IDFT} & \ref{Definitions}\\
$\mathbb{F}_q[\underline{x}]$ & ring of $N$-variable polynomials & \ref{vector spaces}\\
$Z_\Psi$ & an ideal of $\mathbb{F}_q[\underline{x}]$ & \ref{vector spaces}\\
$\preceq$ & a fixed monomial order & \ref{vector spaces}\\
$\mathrm{LM}(f)$ & leading monomial \eqref{LM}
& \ref{vector spaces}\\
$\mathrm{mdeg}\left(\underline{x}^{\underline{d}}\right)$ & $\underline{d}$ for $\underline{x}^{\underline{d}}\in\mathbb{F}_q[\underline{x}]$ & \ref{vector spaces}\\
$\mathrm{ev}$ & evaluation map \eqref{evaluation},\eqref{ev} & \ref{vector spaces}\\
$\mathcal{P}$ & proper transform \eqref{partial DFT} & \ref{vector spaces}\\
$\mathcal{G}_\Psi$ & a Gr\"obner basis of $Z_\Psi$ & \ref{Extension map}\\
$g^{(w)}$ & an element of $\mathcal{G}_\Psi$ & \ref{Extension map}\\
$\mathcal{E}$ & extension map \eqref{E} & \ref{Extension map}\\
$\mathcal{I}$ & inclusion map \eqref{inclusion} & \ref{Extension map}\\
$\underline{a}_w$ & leading monomial $\mathrm{LM}\left(g^{(w)}\right)$ & \ref{Extension map}\\
$\underline{a}+\underline{b}$ & component-wise addition & \ref{Extension map}\\
$\underline{a}\ge\underline{b}$ & component-wise inequality & \ref{Extension map}\\
$\mathcal{R}$ & restriction map \eqref{restriction} & \ref{Isomorphic map}\\
$\mathcal{C}$ & $\mathcal{R}\circ\mathcal{F}^{-1}\circ\mathcal{E}$ in Main Lemma & \ref{Isomorphic map}\\
$U$ & a subspace of $V_{D(\Psi)}$ & \ref{arbitrary subset}\\
$C(U,\Psi)$ & $\mathrm{ev}(U)$ in \eqref{ev image} & \ref{arbitrary subset}\\
$C^\perp(U,\Psi)$ & $\mathrm{ev}(U)^\perp=\mathcal{C}\left(U^\perp\right)$ in \eqref{dual code},\eqref{C image} & \ref{arbitrary subset}\\
$k$ & dimension of $C^\perp(U,\Psi)$ & \ref{arbitrary subset}\\
$B$ & a subset of $D(\Psi)$ & \ref{arbitrary subset}\\
$V_{D\backslash B}$ & $V_B^\perp$ in $V_D$ & \ref{arbitrary subset}\\
$\left(e_{\underline{\psi}}\right)_\Psi$ & an erasure-and-error vector in $V_\Psi$ & \ref{non-systematic case}\\
$\left(r_{\underline{\psi}}\right)_\Psi$ &
a received word $\left(c_{\underline{\psi}}\right)_\Psi+\left(e_{\underline{\psi}}\right)_\Psi$ & \ref{non-systematic case}\\
$\Phi_1$ & set of erasure locations & \ref{non-systematic case}\\
$\Phi_2$ & set of error locations & \ref{non-systematic case}\\
$\Phi$ & set of redundant locations & \ref{Systematic encoding}\\
\hline
\end{tabular}
\end{center}
}
\end{table}

%%%%%%%%%%%%%%%%%%%%%%%%%%%%%%%%%%%%%%%%%%%%%%%%%%%%%%%%%%%%%%%%
\section*{Acknowledgment}
The author would like to thank the anonymous referees for their useful comments that helped improve the final presentation of the paper.

%%%%%%%%%%%%%%%%%%%%%%%%%%%%%%%%%%%%%%%%%%%%%%%%%%%%%%%%%%%%%%%%

\vspace{5mm}
{\bf Hajime Matsui}
received a B.S.\ degree in 1994 from the Department of Mathematics, Shizuoka University, Japan, an M.S.\ degree in 1996 from the Graduate School of Science and Technology, Niigata University, Japan, and Ph.D.\ in 1999 from the Graduate School of Mathematics, Nagoya University, Japan. From 1999 to 2002, he was a Postdoctorate Fellow in the Department of Electronics and Information Science at the Toyota Technological Institute, Japan. From 2002 to 2006, he was a Research Associate at the same institute, where he has been working as an Associate Professor since 2006. His research interests include coding theory, computer science, and number theory.
\end{document}